\documentclass[aps,pra,preprint]{revtex4-1}
\usepackage{graphicx} % for latex2e
\usepackage{amsmath}
\usepackage{amstext}
\usepackage{amsfonts}
\usepackage{amssymb}
\usepackage{color}
\usepackage{longtable}
\usepackage{threeparttable}
\usepackage{rotating}
\usepackage{multirow}

\newcommand{\be}{\begin{equation}}
\newcommand{\ee}{\end{equation}}
\newcommand{\bea}{\begin{eqnarray}}
\newcommand{\eea}{\end{eqnarray}}

\newcommand{\eqa}{\begin{equation}}
\newcommand{\eqz}{\end{equation}}
\newcommand{\eqma}{\begin{eqnarray}}
\newcommand{\eqmz}{\end{eqnarray}}

\newcommand{\Nbas}{N_{\rm bas}}

\newcommand{\dd}{\text{d}}
\newcommand{\mx}[1]{\boldsymbol{#1}}
\newcommand{\bos}[1]{\boldsymbol{#1}}

\usepackage{array}
\newcolumntype{R}[1]{>{\raggedleft  \arraybackslash}p{#1}@{} }
\newcolumntype{C}[1]{>{\centering \arraybackslash}p{#1}@{} }

\newcommand{\pd}[2]{\frac{\partial #1}{\partial #2}}
\newcommand{\cm}{$\text{cm}^{-1}$}
\newcommand{\Vextra}{U}

\newcommand{\Athreed}{A_{\rm 3D}}
\newcommand{\Bthreed}{B_{\rm 3D}}
\newcommand{\Cthreed}{C_{\rm 3D}}
\newcommand{\Fthreed}{F_{\rm 3D}}

\newcommand{\Nquad}{K_{\rm quad}}%{N_{\rm quad}}
\newcommand{\rmSmolyak}{{\rm S}}%{N_{\rm quad}}
\newcommand{\para}[1]{{(#1)}}

\begin{document}

\title{%
Towards breaking the curse of dimensionality in (ro)vibrational computations
of molecular systems with multiple large-amplitude motions
}

\author{Gustavo Avila}
\email{Gustavo$_$Avila@telefonica.net}

\author{Edit M\'atyus}
\email{matyus@chem.elte.hu}

\affiliation{Institute of Chemistry, 
ELTE, E\"otv\"os Lor\'and University
P\'azm\'any P\'eter s\'et\'any 1/A
1117 Budapest, Hungary}

% \date{30 January 2019}
\date{26 March 2019}
\begin{abstract}
\noindent%
Methodological progress is reported in the challenging direction
of a black-box-type variational solution of the (ro)vibrational Schr\"odinger equation
applicable to floppy, polyatomic systems with multiple large-amplitude motions.
%through a combination 
%
This progress is achieved through the combination of
(i) the numerical kinetic-energy operator (KEO) approach 
of 
[E. M\'atyus, G. Czak\'o, and A. G. Cs\'asz\'ar, J. Chem. Phys. 130, 134112 (2009)]
and 
(ii) the Smolyak non-product grid method of 
[G. Avila and T. Carrington, Jr., J. Chem. Phys. 131, 174103 (2009)]. 
The numerical representation of the KEO makes it possible to choose
internal coordinates and a body-fixed frame
best suited for the molecular system. 
The Smolyak scheme reduces the size of the direct-product
grid representation by orders of magnitude, while 
retaining some of the useful features of it.
As a result, multi-dimensional (ro)vibrational states are computed with
system-adapted coordinates, a compact basis- and grid-representation, 
and an iterative eigensolver.
Details of the methodological developments and the first numerical applications 
are presented for the CH$_4\cdot$Ar complex treated in full (12D) vibrational
dimensionality. 
\end{abstract}

\maketitle

%%%%%%%%%%%%%%%%%%%%%%%%%%%%%%%%%%%%%%%%%%%%%%%%%%%%%%%%%%%%%%%%%%%%%%%%%%%%%%%%%%%%%%%
%
% Introduction
%
%%%%%%%%%%%%%%%%%%%%%%%%%%%%%%%%%%%%%%%%%%%%%%%%%%%%%%%%%%%%%%%%%%%%%%%%%%%%%%%%%%%%%%%
\section{Introduction}
\noindent %
Molecular systems with many vibrational degrees of freedom, including 
multiple fluxional modes have been challenging 
for nuclear motion theory (also known as quantum dynamics)
for decades.
These systems are difficult to handle because 
1) they require a curvilinear coordinate representation, 
for which we might not have an analytic kinetic energy operator (KEO) readily available; 
2) their wave functions are spread over multiple wells of the potential energy surface (PES);
and 3) assume the evaluation of high-dimensional (sometimes singular) integrals 
due to the multiple, coupled (curvilinear) internal degrees of freedom.

There are important, high-dimensional molecular systems
with multiple, large-amplitude motions. 
For example, molecular complexes belong to this class. 
Molecular complexes are prototypes for molecular interactions and they can be probed 
in high-resolution spectroscopy experiments. 
Weakly-bound complexes have a shallow PES valley, 
so they exhibit only a few, low-energy transitions between bound states, 
but they usually have a rich predissociation spectrum which can be probed
in overtone spectroscopy experiments. 

The theory of molecular complexes has been
restricted to the explicit quantum mechanical description 
of the inter-monomer modes, while
the monomers were held fixed, described with some rigid, effective structure \cite{HuBrBo08,Avoird:10}.
An explicit consideration of monomer-flexibility effects \cite{TeSu83,ZhWuZhDiBa95}
has come to the focus only in recent years \cite{LeSzAv12,WaCa17,WaCa18WW}. 
This is not surprising: adding the monomer degrees
of freedom to the quantum dynamics treatment rapidly increases the vibrational dimensionality,
while in molecular complexes, monomer flexibility effects are usually small, 
so they can be averaged upon a first look at the system. 
At the same time, the flexibility of monomers, through the kinetic and 
the potential energy couplings, plays a central role in the energy transfer between the inter- and the intra-molecular
degrees of freedom during the (ro)vibrational and collision dynamics. 

Motivation for the present work is provided by these ideas, but we hope that the methodological developments described
in this article will become useful for solving the (ro)vibrational Schr\"odinger equation of (high-dimensional, floppy) molecular systems,
in general.

%%%%%%%%%%%%%%%%%%%%%%%%%%%%%%%%%%%%%%%%%%%%%%%%%%%%%%%%%%%%%%%%%%%%%%%%%%%%%%%%%%%%%%%
% 
% Overview
%
%%%%%%%%%%%%%%%%%%%%%%%%%%%%%%%%%%%%%%%%%%%%%%%%%%%%%%%%%%%%%%%%%%%%%%%%%%%%%%%%%%%%%%%
\section{Curse of dimensionality in vibrational computations \label{ch:curse}}
\noindent %
We focus in the present work on the variational solution of the
Schr\"odinger equation including the (ro)vibrational Hamiltonian of 
$D$ vibrational degrees of freedom, $\boldsymbol{\xi}=(\xi_{1},\ldots,\xi_{D})$, 
\begin{eqnarray}
  \hat{H}_{\rm vib}\Psi_{i}(\xi_{1},\ldots,\xi_{D})
  =
  E_{i} \Psi_{i}(\xi_{1},\ldots,\xi_{D}) \;,
\end{eqnarray}
where the vibrational wave function is as a linear combination of 
$\Nbas$ orthogonal basis functions
\begin{eqnarray}
  \Psi_{i}(\xi_{1},\ldots,\xi_{D})
  =
  \sum_{n=0}^{\Nbas} C^{i}_{n} \psi_{n}(\xi_{1},\ldots,\xi_{D}),
\end{eqnarray}
and the expansion coefficients $C^{i}_{n}$ are obtained as the elements of 
the eigenvectors of the Hamiltonian matrix. The Hamiltonian matrix elements are
computed with some appropriate (multi-dimensional) integration scheme. 
If the basis set is well chosen in this finite basis representation (FBR) scheme, 
the lowest eigenvalues of the Hamiltonian matrix 
converge to the exact energies by increasing $\Nbas$. 
The most common way to build the multi-dimensional basis functions is to
use a direct-product ansatz
\begin{eqnarray}
\psi_{n}(\xi_{1},\ldots,\xi_{D})
= 
\prod_{\chi=1}^{D} 
  \psi^\para{\chi}_{n_{\chi}}(\xi_{\chi})
\end{eqnarray}
constructed from the $\psi^\para{\chi}_{n_\chi}(\xi_\chi)$ orthogonal basis functions. 

\subsection{Curse of dimensionality due to the multi-dimensional vibrational basis}
By adopting a direct-product basis set, the vibrational wave functions are represented as
a linear combination
\begin{eqnarray}
\Psi_{i}(\xi_{1},\ldots,\xi_{D})
=
\sum_{n_{1}=0}^{N_{1}^{\rm max}}
\ldots 
\sum_{n_{D}=0}^{N_{D}^{\rm max}} 
  C^{i}_{n_{1},\ldots,n_{D}}
  \prod_{\chi=1}^{D} 
  \psi^\para{\chi}_{n_{\chi}}(\xi_{\chi})\; , %gab to matyus Psi_{i} is not the same subindex as in Eq 2
\label{expWF}
\end{eqnarray}
in which the number of terms (multi-dimensional basis functions) 
scales exponentially with the vibrational dimensionality, 
$\Nbas=\prod_{\chi=1}^{D}(N_{\chi}^{\rm max}+1)$. 
For low-dimensional systems this is not a problem, but many challenging systems are high dimensional.
Our example system, CH$_{4}\cdot$Ar has twelve vibrational degrees of freedom. 
For a 12-dimensional (12D) problem, if we pick 10 basis functions per coordinate 
(a reasonable starting point if the coordinates are equally coupled), 
the number of product basis functions will be $10^{12}$. In this representation,
we would need to store a vector with $10^{12}$ elements to represent a single vibrational state,
which would require ca. 7.3~TB of memory in double precision arithmetics. 
For this reason, beyond ca.~9 vibrational dimensions, 
it is necessary to develop and use methods which attenuate the curse of dimensionality in
the basis set.

There are different strategies for breaking the exponential growth of the vibrational basis. 
The first option is to improve the quality of the basis functions in order to decrease the number of functions per coordinate, at least for a subset of the coordinates. 
The second option is to find a way to identify and discard the basis functions from the direct-product basis set, which have little effect on the accuracy of the computed eigenvalues.  
The first alternative is efficiently realized by the multi-configuration time-dependent Hartree (MCTDH) method \cite{MCTDH1,MCTDH2}, 
the canonical polyadic (CP) approach \cite{CP1,CP2,MM2} or
in a contracted basis representation obtained by solving reduced-dimensionality eigenproblems \cite{CM1,CM2,CM3}. 
The second alternative is achieved by finding physically motivated restrictions on the basis set indices. 
These restrictions can be as simple as the selection of an appropriate multi-polyad \cite{tc-gab1,tc-gab2}, $f(n_{1},\ldots,n_{D})\le b$,  for which the wave-function expansion reads as
\begin{eqnarray}
  \Psi_{i}(\xi_{1},\ldots,\xi_{D})
  =
  \sum_{f(n_{1},\ldots,n_{D})\le b} 
    C^{i}_{n_{1},\ldots,n_{D}}
    \prod_{\chi=1}^{D} 
    \psi^\para{\chi}_{n_{\chi}}(\xi_{\chi}) \; ,%gab to matyus Psi_{i} is not the same subindex as in Eq 2
\label{exppru}
\end{eqnarray}
This basis-pruning strategy will be used later in this work.
More elaborate basis-pruning restrictions 
are used, for example, in the MULTIMODE program \cite{betterpr1,betterpr2}.

\subsection{Curse of dimensionality due to multi-dimensional integrals}
Reducing the number of the multi-dimensional basis functions 
solves only half of the problem. 
In (ro)vibrational computations, 
multi-dimensional integrals must be evaluated to construct the Hamiltonian matrix.

There are two common ways to cope with the integrals problem. 
The first option is to expand the Hamiltonian in a Sum-of-Products form (SOP).
For example, the potential energy in a SOP form is
\begin{eqnarray}
  \hat{V}(\bos{\xi})
  =
  \sum_{m_{1}=0}^{M_{1}^{\rm max}}
  \ldots 
  \sum_{m_{D}=0}^{M_{D}^{\rm max}} 
    A_{m_{1},\ldots,m_{D}} 
    \prod_{\chi=1}^{D} \mathcal{V}^\para{\chi}_{m_{\chi}}(\xi_{\chi}) \; .
  \label{eq:sop}
\end{eqnarray}
Using the SOP form, multi-dimensional integrals are obtained as 
the sum of products of 1-dimensional (1D) integrals,  
\begin{align}
  \langle %
    \psi^\para{\chi}_{n_{\chi}'} \mid 
    \mathcal{V}^\para{\chi}_{m_{\chi}}\ \mid 
    \psi^\para{\chi}_{n_{\chi}}\rangle 
  &=
  \int 
    \psi^\para{\chi}_{n_{\chi}'}(\xi_{\chi})\ 
    \mathcal{V}^\para{\chi}_{m_{\chi}}(\xi_{\chi}) 
    \psi^\para{\chi}_{n_{\chi}}(\xi_{\chi}) 
    \ \dd \xi_{\chi} \nonumber \\
  &\approx 
  \sum_{k_\chi=1}^{\Nquad} 
    w_{\chi,k_\chi} 
    \psi^\para{\chi}_{n_{\chi}'}(\xi_{\chi,k_{\chi}}) 
    \mathcal{V}^\para{\chi}_{m_{\chi}}(\xi_{\chi,k_{\chi}}) 
    \psi^\para{\chi}_{n_{\chi}}(\xi_{\chi,k_{\chi}}) \; ,
\end{align}
which is evaluated with a 1D numerical quadrature using the
$w_{\chi,k_\chi}$ and $\xi_{\chi,k_{\chi}}$ quadrature weights and points, respectively, 
defined for the $\xi_\chi$ coordinate (in this work, we account for the Jacobian
in the wave function).
The integrals converge to their exact value 
upon the increase of the number of quadrature points, $\Nquad$. 
The SOP form is useful when a small number of terms
is sufficient in Eq.~(\ref{eq:sop}) to represent the Hamiltonian. 
This form is usually employed in MCTDH and in the CP method \cite{MCTDH1,MCTDH2,CP1,CP2,MM2}. 
There are methods which can find an excellent `basis set' for the SOP representation 
of the Hamiltonian \cite{doi:10.1063/1.471513,doi:10.1063/1.471513}.
If the SOP representation, however, requires an excessive number of function evaluations 
over a multi-dimensional grid of the vibrational coordinates,
the exponential scale up with the dimension is re-introduced.
This feature is related to the fact that a SOP representation of the Hamiltonian
can be as expensive as the representation of the multi-dimensional wave function.
In any case, an effective way for attenuating this type of curse of dimensionality was proposed in Ref.~\cite{doi:10.1063/1.471513}.

As an alternative to a sum-of-product representation of the Hamiltonian, 
one can approximate it with a truncated multi-mode expansion of $n$th-order terms \cite{betterpr1,betterpr2,MM2,ZiRa19}. 
For example, a five-mode expansion of the the potential energy is
\begin{align}
V(\xi_{1},\ldots,\xi_{D})
&= 
\sum_{i=1}^{D} V^{i}(\xi_{i}) \nonumber \\
&\quad+ \sum_{i=1}^{D}\sum_{j=i}^{D} V^{i,j}(\xi_{i},\xi_{j}) \nonumber \\
&\quad+ \sum_{i=1}^{D}\sum_{j=i}^{D}\sum_{k=j}^{D} V^{i,j,k}(\xi_{i},\xi_{j},\xi_{k}) \nonumber \\
&\quad+ \sum_{i=1}^{D}\sum_{j=i}^{D}\sum_{k=j}^{D}\sum_{l=k}^{D} V^{i,j,k,l}(\xi_{i},\xi_{j},\xi_{k},\xi_{l}) \nonumber \\
&\quad+ \sum_{i=1}^{D}\sum_{j=i}^{D}\sum_{k=j}^{D}\sum_{l=k}^{D}\sum_{m=l}^{D} V^{i,j,k,l,m}(\xi_{i},\xi_{j},\xi_{k},\xi_{l},\xi_{m}) \; .%gab to matyus there should be restrictions on the subindexes
\end{align}
This expansion is exact if $n=D$, but under certain circumstances (also depending 
on the $\xi_1,\ldots\xi_D$ coordinates) it is very well converged with $n<D$. Using this approximation, the integrals are evaluated using a $D=1,2,\ldots,n$ dimensional direct-product Gauss quadrature, and thereby, the curse of dimensionality is attenuated. 

If we want to use the Hamiltonian directly, without approximating or
expanding it, we have to tackle the direct evaluation of multi-dimensional integrals by multi-dimensional quadrature. 
In this case, the integral of the potential energy over a multi-dimensional basis set is
evaluated as
\begin{eqnarray}
\int \ldots \int
  \prod_{\chi'=1}^{D} 
    \psi^\para{\chi'}_{n_{\chi'}'}(\xi_{\chi'}) 
  V(\xi_{{1}},\ldots,\xi_{{D}})  
  \prod_{\chi=1}^{D} 
  \psi^\para{\chi}_{n_{\chi}}(\xi_{\chi})\ %\mathcal{J}(\xi_{1},\ldots,\xi_{D})\ 
  \dd \xi_{1}\ldots \dd \xi_{D} \nonumber \\ 
\approx
\sum_{K=1}^{\Nquad} 
  w_{K} 
  \prod_{\chi'=1}^{D} 
    \psi^\para{\chi'}_{n_{\chi'}'}(\xi_{\chi',k_{\chi'}}) 
    V(\xi_{1,k_{1}},\ldots,\xi_{D,k_{D}}) 
    \prod_{\chi=1}^{D} 
      \psi^\para{\chi}_{n_{\chi}}(\xi_{\chi,k_{\chi}}) 
    %\mathcal{J}(\xi_{1,k_{1}},\ldots,\xi_{D,k_{D}}) 
    \; , 
\end{eqnarray}
where $w_{k}$ is the multi-dimensional quadrature weight for the $\xi_{\chi,k_\chi}$ points, 
we used the condensed summation index $K\leftrightarrow(k_1,k_2,\ldots,k_D)$. 
The integral approaches its exact value as the $\Nquad$ number of points is increased.
The most common multi-dimensional quadrature is the multi-dimensional direct-product quadrature
\begin{eqnarray}
\int \ldots \int 
  F(\xi_{1},\ldots,\xi_{D})\ 
  \dd \xi_{1}\ldots \dd \xi_{D} 
\approx 
\sum_{k_{1}=1}^{K_{1}^{\rm max}}
\ldots
\sum_{k_{D}=1}^{K_{D}^{\rm max}}  
  w_{1,k_{1}} \ldots w_{D,k_{D}} 
  F(\xi_{1,k_{1}},\ldots,\xi_{D,k_{D}})  \; , 
\end{eqnarray}
where $w_{\chi, k_{\chi}}$ and $q_{\chi,k_{\chi}}$ $(\chi=1,\ldots,D)$ are the 1D
quadrature weights and points for the $\chi$th coordinate. 
1D quadrature rules are most often Gauss (G) quadrature rules,
which integrate exactly 
\begin{eqnarray}
  \int
    w(\xi) F(\xi)\ \dd \xi
  = 
  \sum_{k=1}^{K^{\rm G}_{\rm quad}} 
    w^{\rm G}_k F(\xi_{k}^{\rm G}),
    \quad{\rm for\ any\ }F(\xi)=\sum_{n=0}^{2(K^{\rm G}_{\rm quad}-1)} A_{n}\xi^{n} \; ,
\end{eqnarray}
and $d=2(K^{\rm G}_{\rm quad}-1)$ is called the (1D) accuracy of the Gauss quadrature.

A multi-dimensional direct-product quadrature integration suffers from a similar curse of dimensionality problem as 
a multi-dimensional direct-product basis set: the number of quadrature points, $\Nquad=\prod_{\chi=1}^{D} K_{\chi}^{\rm max}$, 
increases exponentially with the vibrational dimensionality. 
To continue the 12D example from the previous section in which we had 10 basis functions per coordinate,
we choose 13 quadrature points per coordinate (a reasonable value) 
to evaluate the integrals. Then, the number of points in a direct-product grid 
is $2.33\times 10^{13}$. Storage of this many double-precision numbers
would require 170~TB.
%$18.6385$ GB 

As it was explained earlier, the curse of dimensionality in the basis set
can be attenuated by identifying and discarding the product basis functions, which 
are not necessary for the desired precision of the vibrational states. 
Then, we may think about attenuating the curse of dimensionality in the quadrature grid 
by using grids which have a non-product structure.
In general terms, the application of non-product quadrature grids can be justified, 
if the integrand $I(\xi_{1},\ldots,\xi_{D})$ is smooth, 
\emph{i.e.,} it can be expanded with respect to a pruned, product basis set:
\begin{eqnarray}
  I(\xi_{1},\ldots,\xi_{D})
  =
  \sum_{f(n_{1},\ldots,n_{D})\le b} 
    \mathcal{I}_{n_{1},\ldots,n_{D}}
     \prod_{\chi=1}^{D} \Phi^\para{\chi}_{n_{\chi}}(\xi_{\chi})\; .
\end{eqnarray}
For smooth functions, it makes sense to distinguish between \emph{necessary} product basis functions: 
\begin{eqnarray}
  \prod_{\chi=1}^{D} 
    \Phi^\para{\chi}_{n_{\chi}}(\xi_{\chi}) 
  \quad\text{with}~ f(n_{1},\ldots,n_{D})\le b
\end{eqnarray}
and \emph{non-necessary} product basis functions: 
\begin{eqnarray}
  \prod_{\chi=1}^{D} 
  \Phi^\para{\chi}_{n_{\chi}}(\xi_{\chi}) 
  \quad\text{with}~ f(n_{1},\ldots,n_{D})> b \;.
\end{eqnarray}
The total number of necessary and non-necessary product basis functions scales exponentially with the dimension and 
this is the reason why the total number of product quadrature grid points, 
which integrate the overlap of all these functions exactly, also scales exponentially with the dimension.
If we need to integrate accurately only the \emph{necessary} product basis functions, the number of which does not grow exponentially with the dimensionality, it is possible to find a multi-dimensional quadrature, which integrates exactly only the necessary basis functions and which does not grow exponentially with the dimension. In such an approach, the curse of dimensionality in the integration grid 
can be attenuated, \emph{i.e.,}
\begin{align}
&  \int \ldots\int 
  F(\xi_{1},\ldots,\xi_{D})\ \dd \xi_{1}\ldots \dd \xi_{D} 
\approx 
  \sum_{m=1}^{\Nquad} 
    w_{m} F(\xi_{1,m_{1}},\ldots,\xi_{D,m_{D}}) \\
&\quad\quad\text{with}\quad \Nquad\ll \prod_{\chi=1}^{D} K_{\chi}^{\rm max} \; 
\nonumber
\end{align}
during the course of the evaluation of the Hamiltonian terms (without approximating by some expansion).
Optimal non-product quadratures exist for special cases, two of them are explained in the following paragraphs. 

% \vspace{0.3cm}
\subsubsection{An optimal, two-dimensional, non-product quadrature}
The most popular non-product quadrature grid is probably the Lebedev quadrature 
designed to integrate spherical harmonics \cite{lebedev1999quadrature}. Lebedev
grids are used in density functional theory \cite{MuHaLa93}
and they have been used also in rovibrational computations \cite{article123}. 
In particular, if we want to obtain the exact value of all integrals, 
related to the overlap of the spherical harmonics functions,
by numerical integration
\begin{align}
&\int_{0}^{\pi} \dd \theta
\int_{0}^{2\pi} \dd \phi\  
  Y_{l,m}(\theta,\phi) Y_{l',m'}(\theta,\phi) \sin(\theta) \\
&\quad\quad\quad\text{with}\quad l,l'\le l^{\rm max}\quad\text{and}\quad m,m'\le l^{\rm max}, \nonumber
\end{align}
we would need to use a total number of $2(l^{\rm max}+1)^2$ grid points
in the two-dimensional direct-product grid
composed 
of Gauss--Legendre quadrature points for the $\theta$ and Gauss--Chebyshev (first kind)
quadrature points for the $\phi$ coordinate.
Note that in the expansion of the $Y_{l,m}(\theta,\phi) Y_{l',m'}(\theta,\phi) \sin(\theta)$ 
integrand in terms of the product-basis functions, 
one has to comply with the two restrictions, $m \le l$ and $m' \le l'$. 
By taking into account these two restrictions,
a (smaller) non-product quadrature grid, called Lebedev grid, 
can be constructed for the numerical integration which
includes only
\begin{eqnarray}
N_{\rm Leb} \sim \frac{4}{3}(l^{\rm max}+1)^2 
\end{eqnarray}
points, instead of the $2(l^{\rm max}+1)^2$ points of the 2D direct-product grid. 
For example for $l^\text{max}=5$, there are a total number of 36 spherical harmonics functions.
The calculate exactly the overlap of these functions, we would need $2(5+1)^2=72$ points in 
the 2D direct-product grid, whereas it is sufficient to use 50 ($\sim 4/3\cdot(5+1)^2 = 48$)
Lebedev points \cite{lebedev1999quadrature}.
Note that there is not any general formula for the Lebedev quadrature,
but the weights and points are tabulated 
for several two-dimensional maximum accuracy values.

\subsubsection{An optimal, three-dimensional, non-product quadrature}
Our next example is about the calculation of the exact
value of the overlap integrals in a numerical integration scheme,
for products of harmonic oscillator functions,
\begin{align}
&\int_{-\infty}^{\infty} 
\int_{-\infty}^{\infty} 
\int_{-\infty}^{\infty} 
  H_{l}(q_{1})H_{l'}(q_{1}) H_{m}(q_{2})H_{m'}(q_{2}) H_{n}(q_{3})H_{n'}(q_{3})\  
  \text{e}^{-q_{1}^{2}-q_{2}^{2}-q_{3}^{2}}\ \dd q_{1}\ \dd q_{2}\ \dd q_{3} 
\label{stroud}\\
&\quad\quad\quad
\text{with\ the\ restrictions\ } l+n+m\le 4 \text{ and } l'+n'+m'\le 5 \; , 
\label{eq:stroudrest}
\end{align}
where $H_{n}$ is the $n$th Hermite polynomial. 
The smallest, 3D Gauss--Hermite direct-product grid, 
which recovers the exact value for all these integrals 
contains $5^{3}=125$ points.
By explicitly considering the restrictions in Eq.~(\ref{eq:stroudrest}),
we may realize that there are only 35 different product functions in the integrand.
The smallest non-product grid (for a maximum multi-dimensional accuracy of 9), 
which recovers the exact value of the integrals for the possible integrands
consists of only 77 points \cite{stroud71}. We note that the
corresponding Smolyak grid consists of 93 points, which is less than the direct-product grid,
but more than the optimal non-product grid.

In spite of the fact that the optimal multi-dimensional, non-product quadratures use the smallest 
number of points, they have some handicaps. 
First, the construction of optimal, non-product quadratures may be cumbersome.
There are only a limited number of cases for which 
the optimal multi-dimensional quadrature is tabulated in the literature 
(in practice, limited to $D=2$ or 3 for the available cases) \cite{stroud71}: 
the points and weights are available only for certain types of polynomials and for limited values of a maximum multi-dimensional accuracy.
Second, optimal non-product quadratures lack any structure,
which is a serious disadvantage in rovibrational applications \cite{article123}. 
If a non-product grid has some structure (reminiscent of a direct-product grid),
then it can be used to compute sums over the 1D quadrature points sequentially,
which is an important algorithmic element in efficient variational vibrational approaches.

%%%%%%%%%%%%%%%%%%%%%%%%%%%%%%%%%%%%%%%%%%%%%%%%%%%%%%%%%%%%%%%%%%%%%%%%%%%%%%%%%%
\subsubsection{The Smolyak scheme for non-product grids with a structure}
There is a simple way to construct non-product quadrature grids, 
first proposed by the Russian mathematician Sergey A. Smolyak. The Smolyak grid
may be slightly larger than the optimal non-product grid 
but it retains some useful features of direct-product grids. 
The Smolyak scheme was first adopted for
solving the (ro)vibrational Schr\"odinger equation by Avila and Carrington in 2009 \cite{tc-gab1,tc-gab2}
who exploited that the Smolyak grid is built from a sequence of quadrature rules, and its special structure makes it possible to compute the potential and kinetic energy matrix-vector products by doing sums sequentially.
It is possible to combine the Smolyak algorithm with optimal non-product grids of Stroud~\cite{stroud71}, \emph{i.e.,} non-product Smolyak quadrature grids of high-dimensional systems can be constructed from sequences of Stroud-kind non-product quadratures (if
the desired Stroud quadrature is available).
Although the Stroud--Smolyak grids have less structure, they require fewer points than Smolyak quadratures built from 1D quadrature rules.
This direction has been pursued in (ro)vibrational computations 
by Lauvergnat since 2014 \cite{LAUVERGNAT201418}.

%%%%%%%%%%%%%%%%%%%%%%%%%%%%%%%%%%%%%%%%%%%%%%%%%%%%%%%%%%%%%%%%%%%%%%%%%%%%%%%%%%%%%%%%%%%%%%%%%%%%%%%%%
\section{Definition of the (ro)vibrational Hamiltonian in GENIUSH \label{ch:numkeo}}
\noindent The GENIUSH protocol, as it was proposed in 2009 \cite{MaCzCs09}, aimed for 
the development of a universal and exact procedure for the (near-)variational solution
of the (ro)vibrational Schr\"odinger equation.
Its central part is the numerical construction of the 
kinetic energy terms over a grid---thereby, the burdensome derivation and implementation 
of the kinetic energy operator for various molecular and coordinate choices was eliminated.
The GENIUSH program was developed using the discrete variable representation (DVR) \cite{LiCa07},
and it suffered from the curse of dimensionality (Section~\ref{ch:curse}).
The present work aims for the elimination of this bottleneck, both in respect 
of the basis and the grid representations, using the ideas first 
described by Avila and Carrington in 2009 \cite{tc-gab1}.

%%%%%%%%%%%%%%%%%%%%%%%%%%%%%%%%%%%%%%%%%%%%%%%%%%%%%%%%%%%%%%%%%%%%%%%%%%%%%%%%%%%%%%%%%%%%%%%%%%%%%%%%%
\subsection{Numerical representation of the kinetic-energy operator}
The GENIUSH program determines the KEO coefficients numerically, over a grid,
from the user's definition of the vibrational coordinates, $\xi_i\ (i=1,2,\ldots,D)$
(and body-fixed frame definition, which is relevant for rovibrational computations).
Arbitrary coordinates and frames can be defined by writing down
the Cartesian coordinates (in the body-fixed frame) in terms of the vibrational coordinates, $\xi_i\ (i=1,2,\ldots,D)$.
From this coordinate conversion subroutine (written by the user if not yet available in the code), the program numerically evaluates the mass-weighted metric tensor, 
$\mx{g}\in\mathbb{R}^{(D+3)\times(D+3)}$, from the vibrational and the rotational $\mx{t}$ vectors over the coordinate grid. The vibrational $\mx{t}$ vectors are obtained by two-sided finite differences,
for which a step size of $10^{-5}-10^{-7}$~atomic units has been used. 
In principle, the numerical but exact differentiation scheme of Yachmenev and Yurchenko \cite{YaYu15} (using chain rule sequences
and the derivatives of `all' possible elementary functions and thereby
extending Ref.~\cite{LaNa02}) could also be used to eliminate the numerical differentiation step.

The  $\mx{G}$ matrix is calculated by inverting 
$\mx{g}$, $\mx{G}=\mx{g}^{-1}\in\mathbb{R}^{(D+3)\times(D+3)}$, 
over the grid points of the vibrational coordinates. In this notation the last three rows and columns of $\mx{g}$ and $\mx{G}$ 
correspond to the rotational coordinates.
The vibrational kinetic-energy operator has usually been written in the Podolsky form \cite{Pod} 
\begin{eqnarray}
  \hat{T}^{v}_{\rm Pod}
  =
  -\frac{1}{2} 
  \sum_{i=1}^{D} \sum_{j=1}^{D} 
    \tilde{g}^{-1/4} 
    \frac{\partial}{\partial \xi_{i}} 
    G_{i,j} \tilde{g}^{1/2}  
    \frac{\partial}{\partial \xi_{j}} \tilde{g}^{-1/4} 
  \label{eq:pod}
\end{eqnarray}
with $\tilde{g}=\text{det}\mx{g}$, because it requires the calculation of 
only first coordinate derivatives.
The volume element for this Hamiltonian \cite{Pod,MaCzCs09,Ma18nonad}, and for all its rearranged variants, Eqs.~(\ref{eq:rearr}), (\ref{eq:frearr}), (\ref{H12D}) appearing later in this article, is $\dd V= \prod_{i=1}^D\dd \xi_i$.
Ref.~\cite{MaCzCs09} also used a general but ``rearranged''
form of the (ro)vibrational Hamiltonian 
\begin{align}
  \hat{T}^{\rm v}_{\rm rearr}
  &=
  -\frac{1}{2} \sum_{i=1}^{D} \sum_{j=1}^{D} 
    \frac{\partial}{\partial \xi_{i}} 
    G_{i,j} 
    \frac{\partial}{\partial \xi_{j}} 
  +
  \Vextra \label{eq:rearr} \\
  \quad\text{with}\quad
  \Vextra 
  &= 
  \frac{1}{32}\sum_{kl=1}^D \left[\frac{G_{kl}}{\tilde{g}^2}
  \pd{\tilde{g}}{\xi_k} \pd{\tilde{g}}{\xi_l} + 4 
  \pd{}{\xi_k}\left(\frac{G_{kl}}{\tilde{g}}\pd{\tilde{g}}{\xi_l}\right)\right],\label{eq:extralg} \\
  &= 
  \frac{1}{32}\sum_{kl=1}^D \left[\frac{G_{kl}}{\tilde{G}^2}
  \pd{\tilde{G}}{\xi_k} \pd{\tilde{G}}{\xi_l} 
  - 4 \pd{}{\xi_k}\left(\frac{G_{kl}}{\tilde{G}}\pd{\tilde{G}}{\xi_l}\right)\right], \label{eq:extrauG} 
\end{align}
which can be further rearranged to
\begin{align}
  \hat{T}^{\rm v}_{\rm frearr}
  &=
  -\frac{1}{2} \sum_{i=1}^{D} \sum_{j=1}^{D} 
  G_{i,j} 
  \frac{\partial}{\partial \xi_{i}} 
  \frac{\partial}{\partial \xi_{j}} 
  -\frac{1}{2} \sum_{j=1}^{D} B_j
  \frac{\partial}{\partial \xi_{j}}   
  +
  \Vextra \label{eq:frearr} \\
  \text{with}\quad
  B_j
  &=
  \sum_{i=1}^D\frac{\partial}{\partial \xi_{i}} G_{i,j} \; ,
  \label{eq:frearrB}
\end{align}
This last form was used by Lauvergnat and Nauts 
in their numerical KEO approach \cite{LaNa02}.
Eqs.~(\ref{eq:rearr})--(\ref{eq:extrauG}) and 
(\ref{eq:frearr})--(\ref{eq:frearrB}) require third-order derivatives
of the coordinates, which are obtained in GENIUSH by using 
quadruple precision arithmetic
to ensure numerical stability for the finite differences. 
All functions appearing next to the differential
operators in Eqs.~(\ref{eq:pod})--(\ref{eq:frearrB}) have been available from 
the original implementation \cite{MaCzCs09},
so we were able to change between different KEO representations, which has turned out to be necessary for this work \emph{(vide infra)}.

As a first step for implementing the Smolyak algorithm, we had to replace the original DVR implementation with FBR, because
we wanted to discard functions from the direct product using simple, physical arguments, \emph{e.g.,} to restrict the basis to a certain (multi) polyad, Eq.~(\ref{exppru}).

It is important to notice that the application of the Podolsky form, Eq.~(\ref{eq:pod}), assumes the insertion of multiple (truncated)
resolutions of identities in the basis during the construction of the KEO representation. 
In our earlier DVR applications, this did not cause any problem, but since we are aiming for a compact FBR, 
an accurate representation of the Podolsky form could be ensured only if an auxiliary basis set was introduced to converge
the completeness relation
\begin{eqnarray}
\hat{I}\approx\sum_{n=0}^{N_\text{aux}} \mid n \rangle  \langle n \mid.
\end{eqnarray}
For example, in a 3D FBR computation with a basis set
\begin{eqnarray}
\mid n_{1},n_{2},n_{3} \rangle,~ 0 \le n_{1} \le N_{1}^{\rm max},~ 0 \le n_{2} \le N_{2}^{\rm max},~ 0 \le n_{3} \le N_{3}^{\rm max},
\end{eqnarray}
the  matrix-vector products  
\begin{eqnarray}
v^{1}=\tilde{g}^{-1/4} v^{0} \nonumber \\
v^{2}=G_{i,j} \tilde{g}^{1/2}  \frac{\partial}{\partial \xi_{j}} v^{1}
\end{eqnarray}
would have to be expanded with respect to a larger, basis
\begin{align}
&\mid n_{1},n_{2},n_{3} \rangle: \nonumber \\
&\quad 0 \le n_{1} \le N_{1}^{\rm max}+m,
~ 0 \le n_{2} \le N_{2}^{\rm max}+m,~ 
0 \le n_{3} \le N_{3}^{\rm max}+m \; ,
\end{align}
where $m$ is determined by the coordinate-dependence of the $\tilde{g}^{-1/4}$ and $G_{i,j} \tilde{g}^{1/2}$ multi-dimensional functions. 
For the example of the H$_2$O molecule, $m=4$ was found to be sufficient 
to compute the first fifty vibrational states.
So, in this 3D problem, the use of an auxiliary basis set introduces only a modest increase in the computational cost. 
For a 12D problem, however, an $m=4$ choice would increase the basis space by two orders of magnitude!

For this reason, we will use (the rearranged and) the fully rearranged form of
the KEO, Eqs.~(\ref{eq:rearr})--(\ref{eq:frearr}), 
which did not require the introduction of any additional (auxiliary) functions in an FBR computation.
Further details concerning the matrix representation of the KEO, including a pragmatic `treatment' of the KEO singularities, 
ubiquitous in floppy systems, will be explained in Section~\ref{ch:singularity}.

%%%%%%%%%%%%%%%%%%%%%%%%%%%%%%%%%%%%%%%%%%%%%%%%%%%%%%%%%%%%%%%%%%%%%%%%%%%%%%%%%%%%%%%%%%%%%%%%%%%%%%%%%%
\subsubsection{Definition of the coordinates for the example of CH$_4\cdot$Ar \label{ch:coordmetar}}
The vibrational dynamics of the CH$_{4}\cdot$Ar complex was described using
the $\xi_1=R\in[0,+\infty)$, $\xi_2=\theta\in[0,\pi]$, $\xi_3=\phi\in[0,2\pi)$ spherical coordinates, 
and 
the nine dimensionless normal coordinates of the isolated CH$_{4}$ molecule,
$\xi_{3+i}=q_i\in(-\infty,+\infty)$ ($i=1,\ldots,9$). 
At the reference structure (necessary to define the normal coordinates), 
the methane was oriented in the most symmetric fashion
in the Cartesian space with the C atom is at the origin (this orientation also ensured 
that the KEO singularity is not at the equilibrium structure of the complex):
\begin{align}
&\text{H}_{1}:\ \mx{c}^{\rm eq}_1=(r,r,r) \; , \nonumber \\
&\text{H}_{2}:\ \mx{c}^{\rm eq}_2=(r,-r,-r) \; , \nonumber \\ 
&\text{H}_{3}:\ \mx{c}^{\rm eq}_3=(-r,-r,r) \; , \nonumber \\ 
&\text{H}_{4}:\ \mx{c}^{\rm eq}_4=(-r,r,-r) \; , 
\end{align}
and $r=r_{\rm CH}^{\rm eq}=2.052\ 410\ 803$~bohr was the equilibrium C--H distance 
corresponding to the PES of Ref.~\cite{wang2014using}. 
The GENIUSH program evaluates functions appearing in the KEO from
a coordinate conversion routine in which the instantaneous (body-fixed) Cartesian
coordinates must be specified in terms of the internal coordinates.
The Cartesian positions of the carbon and the hydrogen atoms were calculated from 
the $q_1,\ldots,q_9$ normal coordinate values as
\begin{eqnarray}
  \epsilon_{i}=c_{i\epsilon}^{\rm eq}+\sum_{j=1}^{9} l_{\epsilon_{i},j} q_{j}
\end{eqnarray}
where $\epsilon=x,y,z$ and $i=1,2,\ldots 5$, and
the Cartesian coordinates of the Ar atom, $\epsilon_{6}$ $(\epsilon=x,y,z)$, were 
measured from the center of mass of the methane moiety and were obtained as 
\begin{eqnarray}
x_{6}=R \sin\theta \cos\phi \qquad
y_{6}=R \sin\theta \sin\phi \qquad
z_{6}=R \cos\theta  \; .
\end{eqnarray}
In the last step of the calculation of the Cartesian coordinates, 
the center of mass of the complex was shifted to the origin. 
The orientetation of the body-fixed frame corresponding to the coordinates just described 
corresponds to the orientation of the frame used to define the methane's normal coordinates.
A more sophisticated choice of the body-fixed frame can be useful to make rovibrational
computations efficient. In the present work however, we focus on the computation
of the vibrational states.
We used the atomic masses \cite{NIST} $m(\text{H})=1.007\,825\,032\,23$~u, $m(\text{C})=12$~u,
and $m(\text{Ar})=39.962\,383\,123\,7$~u throughout this work.

\subsubsection{Potential energy surface}
Due to the lack of any full-dimensional methane-argon potential energy surface,
we used the sum of the 3D intermolecular potential energy surface of
Ref.~\cite{B009741L,doi:10.1063/1.1506153}
and the 9D methane PES from Wang and Carrington \cite{wang2014using}. 
This setup allows us to study the kinetic coupling of this weakly bound complex. 
Should a full-dimensional PES become available, the computations can be adapted to it.

%%%%%%%%%%%%%%%%%%%%%%%%%%%%%%%%%%%%%%%%%%%%%%%%%%%%%%%%%%%%%%%%%%%%%%%%%%%%%%%%%%%%%%%%%%%
%
% Basis functions, quadrature grid, and matrix vector products
%
%%%%%%%%%%%%%%%%%%%%%%%%%%%%%%%%%%%%%%%%%%%%%%%%%%%%%%%%%%%%%%%%%%%%%%%%%%%%%%%%%%%%%%%%%%%
\section{Implementation of the Smolyak scheme in GENIUSH \label{ch:smolgen}}
\subsection{Pruning the basis functions \label{ch:basispruning}} 
For the example of the CH$_{4}\cdot$Ar complex described with the
$(R,\theta,\phi,q_1,\ldots,q_9)$ vibrational coordinates defined in Section~\ref{ch:coordmetar}, 
we chose the following 1D basis functions:
$\mathcal{L}^{(\alpha)}$ generalized Laguerre basis functions (with $\alpha=2$) or 
tridiagonal Morse basis functions for $R$; 
Legendre basis functions (and variants of them) or Jacobi associated basis functions for $\theta$; 
Fourier functions, composed of $\cos(n_{\phi}\phi),\sin(n_{\phi}\phi)$, for $\phi$;
and harmonic oscillator functions for the $q_{1},\ldots,q_{9}$ methane normal coordinates.
As a result, the direct-product expansion of the vibrational wave function can be written as
\begin{align}
&\Psi_{i}(R,\theta,\phi,q_{1},\ldots,q_{9})
=\nonumber \\
&\sum_{n_{R}=0}^{N_{R}^{\rm max}} 
\sum_{n_{\theta}=0}^{N_{\theta}^{\rm max}}
\sum_{n_{\phi}=0}^{N_{\phi}^{\rm max}}
\sum_{n_{q_{1}}=0}^{b} \ldots \sum_{n_{q_{9}}=0}^{b} %\nonumber \\
  C^{i}_{n_{R},n_{\theta},n_{\phi},n_{q_{1}},\ldots,n_{q_{9}}} %\nonumber \\
 \psi^\para{R}_{n_{R}}(R)
 \psi^\para{\theta}_{n_{\theta}}(\theta)
 \psi^\para{\phi}_{n_{\phi}}(\phi)
 \psi^\para{q_1}_{n_{q_{1}}}(q_{1})\ldots 
 \psi^\para{q_9}_{n_{q_{9}}}(q_{9}) \; .
\label{expWF1}
\end{align}
This direct-product basis representation, for the typical values of 
$N_{R}^{\rm max}>10$, $N_{\theta}^{\rm max}>20$, $N_{\phi}^{\rm max}> 15$, and $b>2$, would include 
%59.049.000 
$>5.90\cdot 10^7$
functions. 
To reduce the basis set size, we prune the basis representation of the methane fragment 
\begin{align}
&\Psi_{i}(R,\theta,\phi,q_{1},\ldots,q_{9})
=\nonumber \\
&\sum_{n_{R}=0}^{N_{R}^{\rm max}} 
\sum_{n_{\theta}=0}^{N_{\theta}^{\rm max}}
\sum_{n_{\phi}=0}^{N_{\phi}^{\rm max}}
\sum_{n_{q_{1}}+\ldots+n_{q_{9}}\le b} %\nonumber \\
C^{i}_{n_{R},n_{\theta},n_{\phi},n_{q_{1}},\ldots,n_{q_{9}}} %\nonumber \\ 
 \psi^\para{R}_{n_{R}}(R)
 \psi^\para{\theta}_{n_{\theta}}(\theta)
 \psi^\para{\phi}_{n_{\phi}}(\phi)
 \psi^\para{q_1}_{n_{q_{1}}}(q_{1})\ldots 
 \psi^\para{q_9}_{n_{q_{9}}}(q_{9}) \; .
\label{expWFpru}
\end{align}
by
replacing the 0 and $b$ lower and upper summation limits of each normal coordinate 
with the basis-pruning condition 
\begin{align}
  n_{q_{1}}+\ldots+n_{q_{9}}\le b\quad (b\in\mathbb{N}_0) \;,
  \label{eq:standardpruning}
\end{align}
which we call `standard' pruning. This condition
is a natural choice for normal coordinates and harmonic oscillator basis functions,
which provide a good `zeroth-order' model.
This standard pruning, equivalent to choosing a big polyad of states, allows us to discard basis functions, for which 
the coupling between the intramolecular basis functions (through the full Hamiltonian) 
is small and for which the zeroth-order energies are very different.
The larger the $b$ value in Eq.~(\ref{eq:standardpruning}), 
the more accurate (higher excited) vibrational states of methane are obtained. 
(If we focused on the computation of highly excited methane states, it would be better to use a more sophisticated
pruning condition.)
For the intermolecular basis set $\psi_{R,n_{R}}(R) \psi_{\theta,n_{\theta}}(\theta) \psi_{\phi,n_{\phi}}(\phi)$
we do not introduce any pruning, because the selected functions are not close to 
any zeroth-order approximate basis set for this system, so we cannot discard any of the functions based on simple arguments. 
Nevertheless, standard pruning of the methane part already reduces the basis set substantially. 
The storage of one vector in the direct-product basis set with 10 basis functions per coordinate would require ca. 8~TB of memory, 
while using standard pruning, Eq.~(\ref{eq:standardpruning}), reduces this value to 0.39~GB.

%%%%%%%%%%%%%%%%%%%%%%%%%%%%%%%%%%%%%%%%%%%%%%%%%%%%%%%%%%%%%%%%%%%%%%%%%%%%%%%%%%%%%%%%%%%%%%%
\subsection{Pruning the grid with the Smolyak scheme \label{ch:gridpruning}} 
The GENIUSH program computes the values of the $G_{i,j}, B_i,$ and $\Vextra$ multi-dimensional
functions of the KEO at multi-dimensional points of the vibrational coordinates. Since we do not use
any interpolation procedure to fit $G_{i,j}, B_i,$ and $V+\Vextra$ to special analytic functions, 
a multi-dimensional quadrature grid is necessary to obtain the integrals.

It is straightforward to design non-product quadrature grids for 
the evaluation of the multi-dimensional integrals of the Hamiltonian operator
with the standard basis-pruning condition, Eq.~(\ref{eq:standardpruning}).
For the example of the CH$_{4}\cdot$Ar complex (see Sec.~\ref{ch:coordmetar} 
for the coordinate definition
and Sec.~\ref{ch:basispruning} for the basis set and the pruning condition), 
the 12D Smolyak integration operator of order $H$ \cite{tc-gab1,tc-gab2}, is 
\begin{align}
  \hat{Q}(12,H) 
  &=
  \sum_{\sigma \mx{g}(\mx{i}) \leq H}
    \otimes\prod_{\chi}^{12}
      \Delta\hat{Q}^{i_\chi}_\chi
  \nonumber \\
  &=
  \sum_{\sigma \mx{g}(\mx{i}) \leq H}
  \Delta \hat{Q}^{i_{R}}_R \otimes \Delta \hat{Q}^{i_{\theta}}_\theta \otimes \Delta \hat{Q}^{i_{\phi}}_\phi  
    \otimes \Delta \hat{Q}^{i_{q_{1}}}_{q_1} \otimes \ldots \otimes \Delta \hat{Q}^{i_{q_{9}}}_{q_9}\; , 
  \label{smolgridor} \\
  &\text{with}\quad i_{\chi}=1,2,3,4,\ldots 
  \quad\text{and}\quad 
  \chi=1(R),2(\theta),\ldots,12(q_9)\; , \nonumber
\end{align}
and the general grid-pruning condition is
\begin{align}
  \sigma \mx{g}(\mx{i}) \leq H: 
  g^{R}(i_{R})+g^{\theta}(i_{\theta})+g^{\phi}(i_{\phi})
  +g^{q_{1}}(i_{q_{1}})+\ldots+g^{q_{9}}(i_{q_{9}})\le H \; .%\nonumber \\  
  \label{eq:gridpruning}
\end{align}
The $i_\chi$th incremental operator is defined as
\begin{eqnarray}
  \Delta \hat{Q}^{i_{\chi}}_\chi = \hat{Q}^{i_{\chi}}_\chi-\hat{Q}^{i_{\chi}-1}_\chi
\end{eqnarray}
with $\hat{Q}^{0}_\chi=0$. 
The action of the $i_\chi$th operator, 
$\hat{Q}^{i_\chi}_\chi$, on an $f$ function is its (numerical, quadrature) integral:
\begin{eqnarray}
  \hat{Q}^{i_{\chi}}_\chi f(\xi_{\chi})
  =
  \sum_{m=1}^{m_{\chi}(i_{\chi})} 
    w^{i_{\chi}}_{\chi,m} f(\xi^{i_{\chi}}_{\chi,m})
\end{eqnarray}
corresponding to the
$w^{i_{\chi}}_{\chi,m}$ quadrature weights and $q^{i_{\chi}}_{\chi,m}$ quadrature points, $m=1,\ldots,m_{\chi}(i_\chi)$
within the $i_\chi$th grid.

We also note that Eq.~(\ref{smolgridor}) can be written
as a linear combination of the 1D integration operators 
(instead of using the incremental operators) as
\begin{align}
  \hat{Q}(12,H)
  &=
  \sum_{\sigma \mx{g}(\mx{i}) \leq H}
    C_{\mx{i}}
    \otimes\prod_{\chi=1}^{12}
      \hat{Q}^{i_{\chi}}_\chi
  \nonumber \\
  &=\sum_{\sigma \mx{g}(\mx{i}) \leq H}
    C_{i_{R},i_{\theta},i_{\phi},i_{q_{1}},\ldots,i_{q_{9}}} 
    \hat{Q}^{i_{R}}_R \otimes \hat{Q}^{i_{\theta}}_\theta \otimes \hat{Q}^{i_{\phi}}_\phi \otimes  
    \hat{Q}^{i_{q_{1}}}_{q_1} \otimes \ldots \otimes \hat{Q}^{i_{q_{9}}}_{q_9} \nonumber \\
  &\quad\text{with}\quad i_{\chi}=1,2,3,4,\ldots 
  \quad\text{and}\quad 
  \chi=1(R),2(\theta),\ldots,12(q_9) \; ,
\end{align}
which allows us to better understand the structure of the Smolyak grid.
The Smolyak quadrature grid is a linear combination of product quadrature grids with different 
1D accuracies, while it has a smaller number of points than the product grid 
$\otimes\prod_{i_\chi=1}^{12}
\hat{Q}^{i^{\rm max}_{\chi}}_\chi
=
\hat{Q}^{i^{\rm max}_{R}}_R \otimes \hat{Q}^{i^{\rm max}_{\theta}}_\theta \otimes 
\hat{Q}^{i^{\rm max}_{\phi}}_\phi \otimes  \hat{Q}^{i^{\rm max}_{q_{1}}}_{q_1} \otimes \ldots \otimes 
\hat{Q}^{i^{\rm max}_{q_{9}}}_{q_9}$, where $i^{\rm max}_\chi=H-\sum_{\chi'\neq \chi} g^{\chi'}(1)$
is determined by the smallest value of the pruning function for the other coordinates, 
Eq.~(\ref{eq:ourgridpruning}).
If a product basis function, 
$
f^\para{R}_{n_{R}}(R) f^\para{\theta}_{n_{\theta}}(\theta) f^\para{\phi}_{n_{\phi}}(\phi) 
f^\para{q_1}_{n_{q_{1}}}(q_{1})\ldots f^\para{q_9}_{n_{q_{9}}}(q_{9})$, 
can be integrated exactly by the product quadrature grid 
$\hat{Q}^{i_{R}}_R \otimes \hat{Q}^{i_{\theta}}_\theta \otimes \hat{Q}^{i_{\phi}}_\phi \otimes  \hat{Q}^{i_{q_{1}}}_{q_1} \otimes \ldots \otimes \hat{Q}^{i_{q_{9}}}_{q_9}$, 
that product basis function is also exactly integrated by the Smolyak quadrature grid $\hat{Q}(12,H)$, because it comprises this smaller product grid \cite{Cools1999}.

To ensure accurate integration, we have to tune three factors: 
a) the pruning function, $g^{\chi}(i_{\chi})$ (which must be a monotonic increasing function);
b) the value of $H$ (the larger, the better); and 
b) the number of grid points, $m_\chi(i_{\chi})$, in the 1D grids
determined by the smallest possible Smolyak grid 
which integrates accurately the Hamiltonian
for a selected, pruned, multi-dimensional basis set. 

For the case of CH$_4\cdot$Ar, the basis-set pruning condition was (Section~\ref{ch:basispruning})
\begin{align}
&0 \le n_{R} \le N^{\rm max}_{R} \nonumber \\
&0 \le n_{\theta} \le N^{\rm max}_{\theta} \nonumber \\
&0 \le n_{\phi} \le N^{\rm max}_{\phi} \nonumber \\
&0 \le n_{q_{1}}+n_{q_{2}}+n_{q_{3}}+n_{q_{4}}+n_{q_{5}}+n_{q_{6}}+n_{q_{7}}+n_{q_{8}}+n_{q_{9}} \le b,
\label{eq:ourbasispruning}
\end{align}
\emph{i.e.,} the intermolecular basis was retained in its product form and 
pruning was introduced for the methane fragment.
The corresponding non-product grid includes the intermolecular grid in its product form, and 
a pruned intramolecular grid implemented using following grid-pruning functions:
\begin{align}
&g^{\chi}(i_{\chi})
=
\left\lbrace
\begin{array}{@{}l l@{}}
  1 \; ,       & \text{for }\chi=R,\theta,\phi  \\
  i \; ,& \text{for }\chi=q_{1},\ldots,q_{9} \;, \quad  i=1,2,3,\ldots  
\end{array}
\right. 
\label{eq:ourgridpruning}
\end{align} 
The corresponding $\hat{Q}^{i_{\chi}}_\chi$ integration operators are chosen as
\begin{align}
&\hat{Q}^{i_{\chi}}_\chi
=
\left\lbrace
\begin{array}{@{}ll@{}}
  \hat{Q}^{M_{\chi}^{\rm max}}_{\chi} \; , & 
     \text{for }\chi=R,\theta,\phi  \\
  \hat{Q}^{m_\chi(i_{\chi})}_{\chi} \; ,& 
    \text{for }\chi=q_{1},\ldots,q_{9} \; , \\   
  & \text{with } m_\chi(i_{\chi})= 1,3,3,7, 9, 9, 9, 9,17,19,19,19,31,33,41,41,\ldots   \\
\end{array}
\right. 
\label{thems}
\\
&\quad\quad\quad\quad\quad\quad\quad\quad\quad\quad
  \text{for\ }i_\chi=1,2,3,\ldots, \ \text{respectively.}  \nonumber
\end{align} 
This choice of the integration operators allowed us
to use the 12D Smolyak operator for the special case 
when the first three degrees of freedom are described with a direct-product grid.
$\hat{Q}^{M_{R}^{\rm max}}_{R}$, 
$\hat{Q}^{M_{\theta}^{\rm max}}_{\theta}$, and 
$\hat{Q}^{M_{\phi}^{\rm max}}_{\phi}$ label the integration operators
corresponding to the spherical degrees of freedom,
and each of them is constructed with a Gauss quadrature rule 
with $M^{\rm max}_\chi$ points and $d_\chi=2M^{\rm max}_\chi-1$ maximum accuracy. 

The $\hat{Q}^{m_\chi(i_{\chi})}_{\chi}\ (\chi=q_1,\ldots,q_9)$ operators, 
corresponding to the normal coordinates, 
are constructed using a nested Hermite quadrature with a maximum degree of 
$d_{\chi}(i_\chi)\geq 2i_\chi -1$, and
$d_{\chi}(i_\chi) =1$, 5, 5, 7, 15, 15, 15, 15, 17, 29, 29, 29, 31, 33, 61, 61$,\ldots$ 
for the $i_\chi=1,2,3\ldots$ sequence of Eq.~(\ref{thems}) 
(also note that the same quadrature is used for each dimensionless normal coordinate).
Nesting means that all quadrature points of the quadrature rule $\hat{Q}^{j}_\chi$ also 
appear in the quadrature rule $\hat{Q}^{j+1}_\chi$. 
It is important that we need to have nested grids to be able to use a Smolyak
quadrature efficiently. 
For this reason, we always use the smallest grid which is nested, \emph{e.g.,}
for $i_\chi=2$ we use a three-point quadrature, $m_\chi(2)=3$, in Eq.~(\ref{thems}), because 
there is not any nested, two-point Hermite quadrature.
Nested Hermite grids are listed in tables, see for example Ref.~\cite{sparseweb}.

In this paragraph, we compare the orders
of magnitudes for a direct-product and 
for a Smolyak grid just defined for the example of CH$_4\cdot$Ar. 
Let us assume, that we have a direct-product basis set with 
$0 \le n_{R},n_{\theta},n_{\phi} \le 9$ functions for the spherical degrees of freedom,
and $0 \le n_{q_{1}}\ldots+n_{q_{9}} \le b=3$ for the methane's degrees of freedom.
The smallest 12D product Gauss grid which gives correctly the overlap integrals for this basis set includes
$10^3 \cdot 4^9 = 2.62\cdot 10^8$ points.
To integrate the overlap for this basis set exactly, we need to choose $H=15$ for the 12D Smolyak grid,
which includes 
$10^3 \times 871 = 8.71\cdot 10^5$ points,
almost three orders of magnitude less than the 12D direct-product Gauss grid. 
Certainly, an even more significant reduction in the grid size (in comparison with a direct-product grid)
can be achieved, if a larger number of degrees of freedom is included in the pruning \cite{AvCa11b}.

The smallest necessary value of $H$ can be calculated 
from the basis-pruning condition and the value of $b$ as follows.
To compute exactly an overlap integral of $2b$ polynomial degree,
it is necessary to have a maximum degree of $2i_\chi -1 \geq 2b$, 
\emph{i.e.,} $i_\chi \geq b + 1/2$.
Then, using the grid-pruning condition, Eq.~(\ref{thems}) and the fact that $i_\chi\geq 1$,
we must have $H\geq b+D$, which makes
$H \geq 3+12 = 15$ for a 12D problem with $b=3$.
In the numerical applications, we choose an $H$ value slightly larger than
this minimal necessary value: $H=b+D+2$ was usually found to be sufficient to converge the results 
for the example computations (Section~\ref{ch:12dim}).

It is important that the Smolyak algorithm uses nested sequences of quadrature rules.
Nesting ensures that the non-product grid has a special structure. 
By exploiting this structure, 
a multi-dimensional integral of a multi-variable function, $F(x_{1},\ldots,x_{D})$, 
can be re-written as
\begin{align}
  &\int \ldots \int F(\xi_{1},\ldots,\xi_{D})
  \ \dd{\xi}_{1}\ldots \dd{\xi}_{D} \nonumber \\
  &\quad =
  \sum_{N=1}^{N^{\rm max}} W_{N} F(\xi_{1,k_{1}},\ldots ,\xi_{D,k_{D}}) 
  \nonumber \\
  &\quad 
  =\sum_{k_{1}=1}^{k_{1}^{\rm max}}\ldots\sum_{k_{D}=1}^{k_{D}^{\rm max}} 
    W^{\rmSmolyak}(k_{1},\ldots,k_{D}) F(\xi_{1,k_{1}},\ldots,\xi_{D,k_{D}}) \;  ,
  \label{eq:multiint}
\end{align}
where the structure of the Smolyak grid appears in the second equation 
through the $k_{i}^{\rm max}$ upper summation indexes \cite{tc-gab1,tc-gab2}:
$k_{1}^{\rm max}$ depends on $H$;
$k_{2}^{\rm max}$ depends on $H$ and $k_{1}^{\rm max}$; 
$k_{3}^{\rm max}$ depends on $H$ and $k_{1}^{\rm max},k_{2}^{\rm max}$, etc.
It is important to notice that 
the multi-dimensional integral, Eq.~(\ref{eq:multiint}), can be written in a sequential sums form (second equation in Eq.~(\ref{eq:multiint}))
only for structured grids, otherwise only the first, computationally more demanding, form is applicable.

\subsection{An efficient matrix-vector product algorithm for computing eigenvalues and eigenvectors with an iterative eigensolver}
We develop a method to compute (ro)vibrational states of polyatomic molecules
with multiple large-amplitude motions.
Probably, the most common way to tackle (ro)vibrational problems is to 
compute the Hamiltonian matrix elements in FBR, and 
then diagonalize the Hamiltonian matrix following the pioneering work
of Whitehead and Handy \cite{WhHa75}.
For polyatomic molecules and complexes, the size of the basis set, 
even if we use a pruned, product basis, may be larger than 100\;000 ($10^5$), and 
a corresponding non-product quadrature grid would consists of more than 10\;000\;000 ($10^7$) points. 
Unless the Hamiltonian matrix is very sparse and the system has 
a high permutation-inversion symmetry, the `traditional' route of using a direct eigensolver 
is not feasible for time and memory reasons.
% }

Using iterative eigensolvers is a practical alternative 
\cite{BrFrWyLe88,BrCa94}, which allows us
to compute eigenvalues and eigenvectors without storing or 
even explicitly computing the Hamiltonian matrix elements. 
The key algorithmic element in relation with iterative eigensolvers, 
is the efficient multiplication of an input vector with the Hamiltonian matrix.

In this section, we develop an efficient matrix-vector products algorithm 
in relation with the numerical KEO approach (Section~\ref{ch:numkeo}) and the Smolyak scheme
(Section~\ref{ch:gridpruning}). 
The multiplication is made efficient by exploiting the structure of the pruned basis set and 
the structure of the non-product Smolyak grid.
Multiplication with the potential energy matrix is carried out as
\begin{align}
v^{\rm out}_{N'_{1\ldots D}}
=& 
\sum_{k_{1}=1}^{k_{1}^{{\rm max}}} 
  T^\para{1}_{n_{1}'}(\xi_{1,k_{1}}) 
\ldots
\sum_{k_{D}=1}^{k_{D}^{{\rm max}}} 
  T^\para{D}_{n_{D}'}(\xi_{D,k_{D}})  \nonumber \\
&\quad \times W^{\rm S}_{K_{1\ldots D}}\ 
V(\boldsymbol{\xi}_{K_{1\ldots D}}) \nonumber \\
&\quad \times \sum_{n_D=0}^{n_D^{\rm max}} T^\para{D}_{n_D}(\xi_{D,k_D})
\ldots
\sum_{n_1=0}^{n_1^{\rm max}} T^\para{1}_{n_1}(\xi_{1,k_1})  \nonumber \\
&\quad \times 
v^{\rm in}_{N_{1\ldots D}} \; 
\label{matvec0} 
\end{align}
with the condensed indexing of the basis labels, grid labels, and multi-dimensional grid points:
\begin{align}
  N_{1\ldots D}\leftrightarrow(n_1,n_2,\ldots,n_D)\;, \quad
  K_{1\ldots D}\leftrightarrow(k_1,k_2,\ldots,k_D)\;, \quad
  \boldsymbol{\xi}_{K_{1\ldots D}}\leftrightarrow(\xi_{1,k_1},\xi_{2,k_2},\ldots,\xi_{D,k_D}), \nonumber
\end{align}
respectively.
$T_{n}(x_{k})$ is the value of the basis function with index $n$ at point $x_{k}$ and 
$W^{{\rmSmolyak}}$ collects the multi-dimensional quadrature weights. 
In the Fortran implementation we use two condensed indexes for the intermediate vectors,
labelled with $K_{1\ldots \Delta}\leftrightarrow(k_1,\ldots,k_\Delta)$ `partial' grid and the corresponding
$N_{\Delta+1,\ldots,D}\leftrightarrow (n_{\Delta+1},\ldots,n_D)$ `partial' basis index. The operations
are performed in parallel using the OpenMP protocol.
The $n^{\rm max}_\chi$ and $k^{\rm max}_\chi$ values for each coordinate, \emph{i.e.,} 
the structure of the basis and the grid, are determined from the basis and the grid pruning conditions. 

For our present numerical example, CH$_4\cdot$Ar, $D=12$ and 
$\boldsymbol{\xi}=(R,c,\phi,q_1,\ldots,q_9)$ (henceforth, we use the short labelling $c=\cos\theta$).
According to the basis pruning condition, Eq.~(\ref{eq:ourbasispruning}),
the upper summation indexes for the basis labels are
\begin{align}
\begin{array}{@{}l@{}}
n_{\chi_i}^{\rm max} 
=
N_{\chi_i}^{\rm max}\;, \quad\text{for } i(\chi_i)=1(R),2(c),3(\phi) \\
n^{\rm max}_{q_{9-i}} 
= 
b - \sum\limits_{j=0}^{i-1} n_{q_{9-j}}\;, \quad \text{for } i=0,1,\ldots,8  \; . \\
\end{array}
\label{eq:basindstruct}
\end{align}
The grid pruning condition in Eq.~(\ref{eq:ourgridpruning}) 
determines the structure of the quadrature indexes according to
\begin{align}
  \begin{array}{@{}l@{}}
    k_{\chi_i}^{\rm max}
    =K_{\chi_i}^{\rm max}\; ,
      \quad\quad\text{for}\quad i(\chi_i)=1(R),2(\theta),3(\phi)\\
    k_{q_i}^{\rm max}
    =m_{q_i}( H-(12-i) - \sum_{j=1}^{i-1} S(k_{q_j}))\; ,
      \quad\quad\text{for}\quad i = 1,2,\ldots, 9 
  \end{array} \; .
  \label{eq:gridindstruct}
\end{align}
where 
$S(k)$ is the index of  the smallest quadrature rule in 
the nested sequence of Hermite quadratures that contains $k$ points.
For the Hermite sequence used in the present work, the $S(k)$ values 
are obtained from Eq.~(\ref{thems}):
\begin{eqnarray}
 S( 1) & = &  1, S( 2)=  2, S( 3)=  2, S( 4)=  4, S( 5)=  4, S( 6)=  4, S( 7)=  4, S( 8)=  5,  \nonumber \\
 S( 9) & = & 5, \ldots, S(17)=  9, S(18)= 10, S(19)= 10, S(20)= 16, \ldots, S(31)= 16,\nonumber \\ 
 S(32) & = & 17, S(33)= 17, S(34)= 18, S(35)= 18, S(36)= 19, S(37)= 19, \nonumber \\ 
 S(38) & = & 20,  S(39)= 20, S(40)= 21, S(41)= 21
\end{eqnarray}
If the FBR method is used for the intermolecular coordinates $R$, $c$, or $\phi$, 
we use more grid points than basis functions $K^{\rm max}>N^{\rm max}$ in order to get exact integrals 
(typically, $K^{\rm max}-N^{\rm max}\approx 5$ was sufficient to achieve convergence). 
If the DVR scheme is used (due to the reasons explained in Section~\ref{ch:singularity}), 
then we have the same number of points and functions, so $K^{\rm max}=N^{\rm max}+1$.

%%%%%%%%%%%%%%%%%%%%%%%%%%%%%%%%%%%%%%%%%%%%%%%%%%%%%%%%%%%%%%%%%%%%%%%%%%%%%%%%%%%%%%%%%%%%%%%%%%%%%%%%%%%%
\subsection{Singularity concerns and a hybrid DVR-FBR solution \label{ch:singularity}}
We have numerically identified that KEO we use to describe the CH$_{4}\cdot$Ar complex 
has singularities along the $\theta$ spherical angle (also related to $\phi$). 
These singularities appear at $\theta=0$ and $\theta=\pi$ ($c=\cos\theta=\pm1$), 
and they represent a considerable challenge for a non-analytic KEO representation, 
especially because two kinds of singularities appear: 
\begin{eqnarray}
\frac{1}{1-c^2}\qquad{\rm and}\qquad\frac{1}{\sqrt{1-c^2}} \; .
\label{eq:singularity}
\end{eqnarray}
This singular property can be discerned from numerical tests 
with the numerical KEO coefficients 
and by calculating matrix elements for the $G_{i,j}\partial^2/\partial\mathcal{R   }_{i}\partial\mathcal{R}_{j}$ terms using an associated Jacobi basis set, $J_{n}^{\alpha,\beta}(c)$, for example.

An obvious way to avoid these types of singular integrals for analytic KEOs would be 
to use the 2D spherical harmonics functions for $\theta$ and $\phi$. 
This option is the way to go for tailor-made approaches, but it would destroy the simplicity and generality 
of a universal (ro)vibrational approach we are developing, especially if there are several groups of 
spherical coordinates $(\theta_i,\phi_i)$ $i=1,2,\ldots$ in the system  \cite{article123}. 
In particular, the application of spherical harmonics would require the development of 
special matrix-vector product routines for each $i=1,2,\ldots$ values.

Another possibility would be to use Jacobi associated functions, $J_{n}^{\alpha,\beta}(c)$, 
with $\alpha$ and $\beta$ close to zero. We could follow this alternative, 
if an analytic KEO and analytic KEO integrals were available.

Since we develop a universal method for numerical KEO representations, we need to find a multi-dimensional quadrature which allows us to evaluate all the different kinds of integrals appearing in the KEO without knowing its exact, analytic form, but knowing only the characteristic
singular behavior, Eq.~(\ref{eq:singularity}).
Let us use Jacobi associated functions with $\alpha=\beta=0.001$ for $c$.
Then, we have to find a quadrature rule which integrates exactly the following types of integrals simultaneously
\begin{eqnarray}
\int_{-1}^{1} J_{n'}^{\alpha,\beta}(c) \frac{1}{1-c^2}J_{n}^{\alpha,\beta}(c)\ \dd c \nonumber \\
\int_{-1}^{1} J_{n'}^{\alpha,\beta}(c) \frac{1}{\sqrt{1-c^2}}J_{n}^{\alpha,\beta}(c)\ \dd c \nonumber \\
\int_{-1}^{1} J_{n'}^{\alpha,\beta}(c) J_{n}^{\alpha,\beta}(c)\ \dd c \label{3in} 
\label{the2sigularities}
\end{eqnarray}
with $0 \le n,n' \le N$. Gauss-quadrature rules exist for each integral 
in Eq.~(\ref{the2sigularities}) separately, but there is not any single Gauss quadrature that integrates exactly 
all three types of integrals, whereas in the numerical KEO, it is not possible to separate different terms 
of different singular behavior (which we know again from numerical test calculations).
Then, the next logical step is to find a (non-Gauss) quadrature rule 
of $M$ points that gives exactly all the integrals in Eq.~(\ref{3in}) at the same time. 
We determined such a quadrature using a two-step procedure.
First, we optimized the quadrature points with a simplex algorithm and calculated the quadrature weights by solving an overdetermined set of equations; this set of points and weights was refined by optimizing both the quadrature points and weights with the simplex algorithm. 
Unfortunately, this (non-Gauss) quadrature includes a large number of points $(K \gg N)$ (three times as many as a single
Gauss-quadrature rule) and some of the points come extremely close to the singular points at $c=-1$ and $c=1$. 
Since GENIUSH calculates the $G_{i,j}$ elements through finite differences, 
the finite step size will place limitations on increasing the number of quadrature points. 
Due to the large number of points and their accumulation near the singular values, 
we cannot accept this special quadrature as a practical solution for the problem,  
but we will use this (non-Gauss) quadrature rule to check the practical ideas we explain in the following paragraphs.

Since we do not have analytic integral expressions, and it is not possible to 
find any compact (Gauss) numerical integration scheme which ensures exact integration,
let us consider approximate integrals (which become accurate at the limit of a large number of points).
First of all, non-exact integration, due to the singularities, 
manifests itself in a non-symmetric matrix representation of the KEO in Eq.~(\ref{eq:frearr}).
Then, instead of aiming for exact integrals (with a compact grid), 
let's aim for a symmetric matrix representation at the first place.
Construction of a symmetric matrix representation is straightforward by 
using Legendre-DVR (or the variants of it discussed below) 
and the inherently more symmetric general KEO in Eq.~(\ref{eq:rearr})
for $c=\cos(\theta)$. We will ensure a symmetric representation in the same way, 
as in the original DVR-based GENIUSH implementation \cite{MaCzCs09}
(see also Ref.~\cite{WeCa94} concerning the Legendre polynomials) and in its applications to floppy systems \cite{FaCsCz13,MaSziCs14,FaSaCs14,SaCs16,SaCsAlWaMa16,SaCsMa17},
which did not suffer from the present singularity problems but which did suffer from the curse of dimensionality.
So, we handle the singular coordinate $c$ as we would do it in GENIUSH-DVR, for the rest of the coordinates, we use FBR.
  
So, instead of using the fully rearranged KEO, Eq.~(\ref{eq:frearr}), 
for which we obtain a non-symmetric matrix representation due to inexact integration
(the off-diagonal elements with different basis indexes of $c$ fail to be equal unless
they are exactly integrated), we re-write the KEO for the $c$ coordinate into the more symmetric form 
\begin{align}
  \hat{T}^{\rm v}
  &=
  -\frac{1}{2} \sum_{j=1}^{12} 
    \frac{\partial}{\partial c} G_{c,j} \frac{\partial}{\partial \xi_{j}} 
  -\frac{1}{2} \sum_{i=1\ne 2}^{12} \sum_{j=1}^{12} 
      G_{i,j} \frac{\partial}{\partial \xi_{i}}\frac{\partial}{\partial \xi_{j}}
  -\frac{1}{2} \sum_{i=1}^{12} B_{i} \frac{\partial}{\partial \xi_{i}}
  + \Vextra, 
\label{H12D} \\
B_{i}&= \sum_{k=1,\ne 2}^{12} \frac{\partial}{\partial \xi_{k}}G_{k,i} \; . \nonumber
\end{align}
Using this KEO and a hybrid DVR(c)-FBR representation, the Hamiltonian matrix
is real, symmetric by construction and
the matrix elements for functions with the same $c$ index are 
the same as the ones we get using the fully rearranged KEO, Eq.~(\ref{eq:frearr}).
We have carried out an additional test for this hybrid DVR-FBR approach.
First, we performed a fully FBR computation with the fully rearranged KEO, Eq.~(\ref{eq:frearr}),
using a Jacobi associated basis set for $c$ with $\alpha=0.01$, $\beta=0.01$ and a (non-Gauss) quadrature developed to calculate accurately the integrals of Eq.~(\ref{the2sigularities}). This non-Gauss quadrature 
included a very large number of points for $c$, 
so we could afford only a small basis and grid for the other degrees
of freedom. We repeated the computation using the same, small basis set for the non-$c$ coordinates and DVR with the symmetric KEO, Eq.~(\ref{H12D}), for $c$. The two computations resulted in the same eigenvalues, which provides a numerical test for our practical DVR-FBR approach (of course, the eigenvalues obtained in this way were different from the converged values due to the smallness of the non-$c$ basis set). 
So, in this sense, using DVR(c)-FBR and the KEO in Eq.~(\ref{H12D}) has the correct ``limiting'' (convergence) behavior, while it ensures a symmetric matrix representation by construction.

%%%%%%%%%%%%%%%%%%%%%%%%%%%%%%%%%%%%%%%%%%%%%%%%%%%%%%%%%%%%%%%%%%%%%%%%%%%%%%%%%%%%%%%%%%%%%%%%%%%%%%%%%%%%
\subsection{Matrix-vector products in the hybrid DVR-FBR \label{eq:mvecprod}}
The matrix-vectors products in the hybrid DVR-FBR 
are carried out similarly to Eq.~(\ref{matvec0}).
In what follows we list the necessary changes in comparison with the 
fully FBR PES multiplication, Eq.~(\ref{matvec0}) 
to accommodate the hybrid FBR-DVR representation for the KEO of Eq.~(\ref{H12D}).
We also note that in the hybrid DVR-FBR scheme, the 
$W^{{\rmSmolyak}}_{K_{1\ldots D}}$
Smolyak weights were obtained using a quadrature rule for the $c$ coordinate with weights equal to one. 

%%%%%%%%%%%%%%%%%%%%%%%%%%%%%%%%%%%%%%%%%%%%%%%%%%%%%%%%%%%%%%
%%%%%%%%%%%%%%%%%%%%%%%%%%%%%%%%%%%%%%%%%%%%%%%%%%%%%%%%%%%%%%
%%%%%%%%%%%%%%%%%%%%%%%%%%%%%%%%%%%%%%%%%%%%%%%%%%%%%%%%%%%%%%
\begin{enumerate}
\item
The matrix-vector product for the potential (and the $U$ pseudo-potential) term
is carried out as in Eq.~(\ref{matvec0}), but
for the $c$ coordinate, we make the following replacements:
\begin{align}
  T_{n'_c}^{(c)}(\xi_{c,k_c}) \rightarrow \delta_{n'_c,k_c-1}
  \quad \text{and} \quad
  T_{n_c}^{(c)}(\xi_{c,k_c}) \rightarrow \delta_{n_c,k_c-1} \; .
\end{align}
%
%%%%%%%%%%%%%%%%%%%%%%%%%%%%%%%%%%%%%%%%%%%%%%%%%%%%%%%%%%%%%%
\item 
The matrix-vector product for the $\frac{\partial}{\partial c}  G_{c,c} \frac{\partial}{\partial c}$ 
term is calculated as in Eq.~(\ref{matvec0}), but for the $c$ coordinate, we make 
the following replacements:
\begin{align}
&V(\boldsymbol{\xi}_{K_{1\ldots D}})
\rightarrow 
G_{c,c}(\boldsymbol{\xi}_{K_{1\ldots D}})\; , \nonumber  \\
&T_{n'_c}^{(c)}(\xi_{c,k_c})   \rightarrow  -M_{k_{c}-1,n_{c}'} \; ,
\qquad\text{and}\qquad
T_{n_c}^{(c)}(\xi_{c,k_c})   \rightarrow  M_{k_{c}-1,n_{c}}
\end{align}
with
\begin{eqnarray}
 M_{n'_{c},n_{c}}=\int_{-1}^{1} \Theta_{n_{c}'}(c) \frac{\dd}{\dd c} \Theta_{n_{c}}(c)\ \dd c \; ,
\end{eqnarray}
where $\Theta_{n_{c}}(c)$ is the $n_c$th (cot-, sincot-)Legendre-DVR function with 
$K_{c}^{\rm max}=N_{c}^{\rm max}+1$ quadrature points \emph{(vide infra)}. 
%
%%%%%%%%%%%%%%%%%%%%%%%%%%%%%%%%%%%%%%%%%%%%%%%%%%%%%%%%%%%%%%
\item
The matrix-vector product for the $\frac{\partial}{\partial c}  G_{c,\mathcal{R}} \frac{\partial}{\partial \mathcal{R}}$ term, where $\mathcal{R}$ is not the $c$ coordinate, is calculated as in Eq.~(\ref{matvec0}) with the following replacements:
\begin{align}
  &V(\boldsymbol{\xi}_{K_{1\ldots D}})
  \rightarrow
  G_{c,\mathcal{R}}(\boldsymbol{\xi}_{K_{1\ldots D}})\; , \nonumber \\
  &T^\para{c}_{n'_c}(\xi_{c,k_c}) \rightarrow -M_{k_{c}-1,n_{c}'} 
  \quad \text{and} \quad
  T^\para{c}_{n_c}(\xi_{c,k_c}) \rightarrow \delta_{n_c,k_c-1} \; , \nonumber \\
  &
  T^\para{\mathcal{R}}_{n_{\mathcal{R}}}(\xi_{R,k_{\mathcal{R}}})
  \rightarrow  
  \frac{d}{d\mathcal{R}} T^\para{\mathcal{R}}_{n_{\mathcal{R}}}(\mathcal{R}) \Big |_{\mathcal{R}=\xi_{\mathcal{R},{k_{\mathcal{R}}}}} \; .
\end{align}
%
%%%%%%%%%%%%%%%%%%%%%%%%%%%%%%%%%%%%%%%%%%%%%%%%%%%%%%%%%%%%%%
\item
The matrix-vector product for the $\frac{\partial}{\partial \mathcal{R}} G_{\mathcal{R},c} \frac{\partial}{\partial c}$ term, where $\mathcal{R}$ is not the $c$ coordinate, is calculated as in Eq.~(\ref{matvec0}) with the following replacements:
\begin{align}
&V(\boldsymbol{\xi}_{K_{1\ldots D}})
\rightarrow 
G_{\mathcal{R},c}(\boldsymbol{\xi}_{K_{1\ldots D}})\; , \nonumber \\
  & T_{n'_c}^{(c)}(\xi_{c,k_c}) \rightarrow \delta_{n'_c,k_c-1}
  \quad \text{and} \quad
  T_{n_c}^{(c)}(\xi_{c,k_c}) \rightarrow M_{k_{c}-1,n_{c}} \; , \nonumber \\
&
T^\para{\mathcal{R}}_{n'_{\mathcal{R}}}(\xi_{\mathcal{R},k_{\mathcal{R}}})   \rightarrow  \frac{d}{d\mathcal{R}} T^\para{\mathcal{R}}_{n_{\mathcal{R}}'}(\mathcal{R}) \Big |_{\mathcal{R}=\xi_{\mathcal{R},{k_{\mathcal{R}}}}} \; .
\end{align}
%    
%%%%%%%%%%%%%%%%%%%%%%%%%%%%%%%%%%%%%%%%%%%%%%%%%%%%%%%%%%%%%%
\item
The matrix-vector product for the $G_{\mathcal{R},\mathcal{R}} \frac{\partial^2}{\partial \mathcal{R}^2}$ term, where $\mathcal{R}$ is not the $c$ coordinate, is calculated as in Eq.~(\ref{matvec0}) with the following replacements:
\begin{align}
&V(\boldsymbol{\xi}_{K_{1\ldots D}})
\rightarrow 
G_{\mathcal{R},\mathcal{R}}(\boldsymbol{\xi}_{K_{1\ldots D}}) \; , \nonumber \\
  & T_{n'_c}^{(c)}(\xi_{c,k_c}) \rightarrow \delta_{n'_c,k_c-1}
  \quad \text{and} \quad
  T_{n_c}^{(c)}(\xi_{c,k_c}) \rightarrow \delta_{n_c,k_c-1} \; , \nonumber \\
&T^\para{\mathcal{R}}_{n_{\mathcal{R}}}(\xi_{\mathcal{R},k_{\mathcal{R}}})   
\rightarrow  
\frac{d^2}{d\mathcal{R}^2} T^\para{\mathcal{R}}_{n_{\mathcal{R}}}(\mathcal{R}) \Big |_{\mathcal{R}=\xi_{\mathcal{R},{k_{\mathcal{R}}}}} \; .
\end{align}
%
%%%%%%%%%%%%%%%%%%%%%%%%%%%%%%%%%%%%%%%%%%%%%%%%%%%%%%%%%%%%%%
\item
The matrix-vector product for the $B_{\mathcal{R}} \frac{\partial}{\partial \mathcal{R}}$ term, where $\mathcal{R}$ is not the $c$ coordinate, is calculated as in Eq.~(\ref{matvec0}) with the following changes
\begin{align}
&V(\boldsymbol{\xi}_{K_{1\ldots D}})
\rightarrow 
B_{\mathcal{R}}(\boldsymbol{\xi}_{K_{1\ldots D}}) \nonumber \; , \nonumber \\
  & T_{n'_c}^{(c)}(\xi_{c,k_c}) \rightarrow \delta_{n'_c,k_c-1}
  \quad \text{and} \quad
  T_{n_c}^{(c)}(\xi_{c,k_c}) \rightarrow \delta_{n_c,k_c-1} \; , \nonumber \\
&T^\para{\mathcal{R}}_{n_{\mathcal{R}}}(\xi_{\mathcal{R},k_{\mathcal{R}}})   \rightarrow  
\frac{d}{d\mathcal{R}} T^\para{\mathcal{R}}_{n_{\mathcal{R}}}(\mathcal{R}) \Big |_{\mathcal{R}=\xi_{\mathcal{R},{k_{\mathcal{R}}}}} \; .
\end{align}
%
%%%%%%%%%%%%%%%%%%%%%%%%%%%%%%%%%%%%%%%%%%%%%%%%%%%%%%%%%%%%%%
\item
The matrix-vector product for the $G_{\mathcal{R}_s,\mathcal{R}_t} \frac{\partial^2}{\partial \mathcal{R}^2}$ term, where $\mathcal{R}_s$ and $\mathcal{R}_t$ are not the $c$ coordinate, is calculated as in Eq.~(\ref{matvec0}) with the replacements
\begin{align}
&V(\boldsymbol{\xi}_{K_{1\ldots D}})
\rightarrow 
G_{\mathcal{R}_s,\mathcal{R}_t}(\boldsymbol{\xi}_{K_{1\ldots D}}) \; , \nonumber \\
  & T_{n'_c}^{(c)}(\xi_{c,k_c}) \rightarrow \delta_{n'_c,k_c-1}
  \quad \text{and} \quad
  T_{n_c}^{(c)}(\xi_{c,k_c}) \rightarrow \delta_{n_c,k_c-1} \; ,\nonumber \\
& T^\para{\mathcal{R}_s}_{n_{\mathcal{R}_s}}(\xi_{\mathcal{R}_s,k_{\mathcal{R}_s}})   
\rightarrow  
\frac{d}{d\mathcal{R}_s} T^\para{\mathcal{R}_s}_{n_{\mathcal{R}_s}}(\mathcal{R}) \Big |_{\mathcal{R}_s=\xi_{\mathcal{R}_s,{k_{\mathcal{R}_s}}}} 
  \quad \text{and} \quad
 T^\para{\mathcal{R}_t}_{n_{\mathcal{R}_t}}(\xi_{\mathcal{R}_t,k_{\mathcal{R}_t}})   
 \rightarrow  
 \frac{d}{d\mathcal{R}_t} T^\para{\mathcal{R}_t}_{n_{\mathcal{R}_t}}(\mathcal{R}) \Big |_{\mathcal{R}_t=\xi_{\mathcal{R}_t,k_{\mathcal{R}_t}}} \; .
\end{align}
\end{enumerate}

%%%%%%%%%%%%%%%%%%%%%%%%%%%%%%%%%%%%%%%%%%%%%%%%%%%%%%%%%%%%%%%%%%%%%%%%%%%%%%%%%%%%%%%%%%%%%%%%%%%%%%%%%%%%
\subsection{Analysis and improvements for the intermolecular representation \label{ch:inter}}
To test the convergence properties, and to determine the optimal basis set and grid sizes for our example system, 
CH$_4\cdot$Ar, we performed reduced-dimensionality computations.
Intermolecular (3D) computations were performed
with a fixed methane structure corresponding to the effective rotational constant, $B_{v=0}=5.246\,981\,98\ \text{cm}^{-1}$ 
(and an effective C--H distance of 
$\langle R(\text{CH})\rangle_{v=0}=1.107\,117\,44$~bohr)
obtained with the ground-state vibrational wavefunction of CH$_{4}$ with pruning condition $b=8$ (see Section~\ref{ch:intra}) and using the isolated methane's PES \cite{wang2014using}.

%%%%%%%%%%%%%%%%%%%%%%%%%%%%%%%%%%%%%%%%%%%%%%%%%%%%%%%%%%%%%%%%%%%%%%%%%%%%%%%%%%%%%%%%%%%%%%%%%%%%%%%%%%%%
\subsubsection{Intermolecular angular representation: Legendre, cot-Legendre and sincot-Legendre DVRs} 
\noindent % 
Since regions near the singularities, Eq.~(\ref{eq:singularity}), are dynamically relevant
for the CH$_{4}\cdot$Ar complex, using Legendre DVR for the coordinate $c=\cos\theta$ is an inefficient choice:
more than 120 points are needed to converge all  vibrational bound states of CH$_4\cdot$Ar~(3D) within 0.01~cm$^{-1}$.

In 2010, Schiffel and Manthe \cite{SCHIFFEL2010118} proposed more efficient alternatives to Legendre DVR
to be used for the type of singularities we have to tackle.
First of all, the quadrature is improved by selecting the quadrature points, different from 
the Legendre points, as the inverse cotangent of the eigenvalues ($w_i$) of the following matrix
\begin{align}
  P_{n,m}
  &=
  \int_{-1}^{1} 
    L_{n}(c)  \frac{c}{1-c^2} L_{m}(c)\ \dd{c}, 
  \quad
  n,m = 0,\ldots,N_{c}^{\rm max}-1 \nonumber \\
  \epsilon_{i}&=\text{arccot}(w_{i}),\quad i=1,\ldots,N_{c}^{\rm max}
\end{align}
where $L_{n}(c)$ is the $n$th normalized Legendre function.  %I DIDN'T IMPLEMENT THIS CHANGE: gab to matyus  L_{n}(c) to  L_{n_{c}}(c) 
These integrals are calculated exactly using the Gauss--Chebyshev quadrature with 
a sufficiently large number of points. Using the eigenvectors, $\mx{A}$, of $\mx{P}$ 
the cot-Legendre DVR basis functions are defined as
\begin{eqnarray}
  \Theta_{n}(c)
  =
  \sum_{m=0}^{N_{c}-1} 
  A_{m,n} L_{m}(c), 
\quad n=0,\ldots,N_{c}^{\rm max}-1 \; ,
 \end{eqnarray}
and the first derivative matrix, $\mx{M}$, for the cot-Legendre DVR functions is
\begin{eqnarray}
  M_{n',n}=\int_{-1}^{1} \Theta_{n'}(c) \frac{\dd}{\dd c} \Theta_{n}(c)\ \dd c \;.  
\end{eqnarray}
In our test calculations, it was sufficient to use 50 cot-Legendre DVR points
to converge all bound states of the CH$_{4}\cdot $Ar in 3D (within 0.01 cm$^{-1}$) (see also Table~\ref{1table}). 

Schiffel and Manthe \cite{SCHIFFEL2010118} continued and proposed further improvements by extending the 
basis set. They have noticed that some eigenfunctions of the KEO in spherical coordinates have a 
$\sin(\theta)$ `component' close to the singularities, so they extended the Legendre basis set 
with sine functions. Their new basis set included
$L_{n}(c)$, $n=0,\ldots,N_{c}^{\rm max}-s$ and 
$\sin(\theta), \ldots, \sin(s\theta)$, where $s=2$ was sufficient 
(and stable without any over-completeness problems, which would occur for larger $s$ values)
in most applications. 
A corresponding DVR basis set, called `sincot-Legendre DVR basis', is obtained in the following procedure:
\begin{enumerate}
  \item 
    Orthogonal basis functions are created from the set 
    $\lbrace L_{n}(x), (n=0,\ldots,N_{c}^{\rm max}-2),\sin\theta,\sin2\theta \rbrace$  %gab to matyus  L_{n}(c) to  L_{n_{c}}(c) 
    by diagonalizing the corresponding overlap matrix $S^{\rm sin-cos}$. 
    The orthogonal basis functions, $L_n^{\rm sin-cos}(c)$,
    are calculated using the eigenvectors of the overlap matrix.
  \item
    A $\mx{P}^{\rm sin-cos}$ matrix is introduced with the elements
    \begin{align}
      P_{n,m}^{\rm sin-cos}
      &=
      \int_{-1}^{1} L_{n}^{\rm sin-cos}(c)  
      \frac{c}{1-c^2} L_{m}^{\rm sin-cos}(c)\ \dd{c},\nonumber \\
      &\quad\quad\quad\quad\quad n,m = 0,\ldots, N_{c}^{\rm max} \; .
    \end{align}
     The DVR points are the inverse cotangent of the $w_i$ eigenvalues of $\mx{P}^{\rm sin-cos}$. 
     The sincot-Legendre DVR basis functions are obtained from the eigenvectors of 
     the $\mx{P}^{\rm sin-cos}$ matrix, collected in $\mx{A}^{\rm sin-cos}$, as
     \begin{eqnarray}
       \Theta^{\rm sin-cos}_{n}(c)
       =
       \sum_{m=0}^{N_{c}^{\rm max}} 
         A_{m,n}^{\rm sin-cos} L_{m}^{\rm sin-cos}(c), 
       \quad n=0,\ldots,N_{c}^{\rm max} \; .
     \end{eqnarray}
   \item
     The first derivative matrix, $\mx{M}^{\rm sin-cos}$, for sincot-Legendre DVR is 
     \begin{eqnarray}
       M^{\rm sin-cos}_{n',n}
       =
       \int_{-1}^{1} 
         \Theta^{\rm sin-cos}_{n'}(c) 
         \frac{\dd{}}{\dd{c}} \Theta^{\rm sin-cos}_{n}(c)\ \dd{c} \; .
     \end{eqnarray}     
\end{enumerate}
The integrals for the $\mx{S}^{\rm sin-cos}$, $\mx{P}^{\rm sin-cos}$, and $\mx{M}^{\rm sin-cos}$ 
matrices can be calculated analytically using elementary properties of trigonometric functions
and they were tabulated in Ref.~\cite{SCHIFFEL2010118}. 

We used the sincot-Legendre DVR points and the corresponding first derivative matrix elements 
(as an alternative to Legendre DVR)
in the matrix-vector multiplication procedure described in Section~\ref{ch:singularity}. 
Our 3D test computations show that it is sufficient to 
use 21 sincot-Legendre DVR points for coordinate $c$ 
to converge all the bound states within 0.01 cm$^{-1}$ for CH$_{4}\cdot$Ar, 
which is a significant reduction compared to the original Legendre DVR which required more than 120 points. 
The performance of a few different representations for the $c$ coordinate 
is compared in Table~\ref{1table}.
In all computations, we used the $\mathcal{L}^\alpha_n$ generalized 
Laguerre polynomials (with $\alpha=2$) for $R$, 
scaled to the $[2.64,30]$~\AA\ interval,
and Fourier functions for $\phi$. 
The number of points used for the $R,\cos\theta,$ and $\phi$ degrees of freedom in the three test sets of the table is 
\begin{itemize}
  \item
  $\Athreed$: $(K^{\rm max}_R,K^{\rm max}_c,K^{\rm max}_\phi)=(81,101(\text{L}),101)$ using Legendre (L) DVR for $c$
  \item
  $\Bthreed$: $(K^{\rm max}_R,K^{\rm max}_c,K^{\rm max}_\phi)=(61,21(\text{SCL}),17)$ using sincot-Legendre (SCL) DVR for $c$
  \item
  $\Cthreed$: $(K^{\rm max}_R,K^{\rm max}_c,K^{\rm max}_\phi)=(61,31(\text{SCL}),31)$ using sincot-Legendre (SCL) DVR for $c$
\end{itemize}
It is important to observe in Table~\ref{1table} that the vibrational states are not perfectly converged 
even with a very large number (more than 100) of Legendre DVR points. 
On the contrary, almost perfect results are obtained with less than 
30 sincot-Legendre DVR points. Another important observation (relevant for the 12D applications in Section~\ref{ch:12dim}) 
is that we can use fewer Fourier basis functions for $\phi$, than (sincot-Legendre) functions 
for $\theta$ to converge the 3D vibrational energies.

\begin{center}
\begin{longtable}{@{}r ccc @{\ \ \ \ \ } rr @{\ \ \ \ \ }rr@{\ \ \ \ \ }rr@{\ \ \ \ \ }r@{}}
\caption{%
Convergence tests for the bound-state vibrational energies of %the angular basis and grid representation for 
CH$_{4}\cdot$Ar (3D) using spherical coordinates, $(R,\cos\theta,\phi)$.
The vibrational energies, $\tilde\nu$ in cm$^{-1}$ and referenced to the ZPVE,
were computed with GENIUSH-DVR and the PES of Ref.~\cite{B009741L,doi:10.1063/1.1506153}.
The vibrational states are labelled 
with the (approximate) $j$ methane angular momentum quantum number, the $n_R$ radial excitation index, 
and the $\Gamma$ $T_{\rm d}$(M) irrep label.
The $R$ and $\phi$ degrees of freedom are described using 
generalized Laguerre basis functions ($\mathcal{L}^{\alpha=2}_n$) scaled to $[2.64,30]$~\AA,
and Fourier functions defined over the $[0,2\pi)$ interval, respectively.
Legendre or sincot-Legendre DVR is used for $\cos\theta$.
The number of basis functions and grid points is given for each set 
as $(K_R^{\rm max},K_c^{\rm max},K_\phi^{\rm max})$.
The test sets, $i=\Athreed,\Bthreed,$ and $\Cthreed$ 
are compared with the ``final'', benchmark values of $\Fthreed$, 
$\Delta \tilde\nu_i=\tilde\nu_i-\tilde\nu_{\Fthreed}$.
\label{1table} }\\
\hline
\hline
\multicolumn{1}{r}{~~~~~} &
\multicolumn{1}{c}{~} &
\multicolumn{1}{c}{~} &
\multicolumn{1}{c}{~} &
\multicolumn{2}{c}{$\cos\theta$: Legendre-DVR} &
\multicolumn{5}{c}{$\cos\theta$: sincot-Legendre} \\
\cline{5-6} \cline{8-11} \\[-0.7cm]
\multicolumn{1}{r}{~~~~~} &
\multicolumn{1}{c}{~} &
\multicolumn{1}{c}{~} &
\multicolumn{1}{c}{~} &
\multicolumn{2}{c}{$\Athreed:$ (111,111,31)} &
\multicolumn{2}{c}{$\Bthreed:$ (111,21,17)} &
\multicolumn{2}{c}{$\Cthreed:$ (111,31,31)} &
\multicolumn{1}{c}{$\Fthreed:$ (151,31,31)} \\
%\hline
%
\multicolumn{1}{c}{$n$} &
\multicolumn{1}{c}{$j$} &
\multicolumn{1}{c}{$n_{R}$} &
\multicolumn{1}{c}{$\Gamma$} &
\multicolumn{1}{c}{$\tilde\nu_{\Athreed}$}&
\multicolumn{1}{c}{$\Delta\tilde\nu_{\Athreed}$}  &
\multicolumn{1}{c}{$\tilde\nu_{\Bthreed}$} &
\multicolumn{1}{l}{$\Delta\tilde\nu_{\Bthreed}$}  &
\multicolumn{1}{c}{$\tilde\nu_{\Cthreed}$} &
\multicolumn{1}{c}{$\Delta\tilde\nu_{\Cthreed}$}  &
\multicolumn{1}{r}{$\tilde\nu_{\Fthreed}$}  \\
\hline
\endfirsthead
\multicolumn{8}{c}{{\tablename} \thetable{}  Continued} \\[0.5ex]
\multicolumn{1}{c}{$n$} &
\multicolumn{1}{c}{$j$} &
\multicolumn{1}{c}{$n_{R}$} &
\multicolumn{1}{c}{$\Gamma$} &
\multicolumn{1}{c}{$\Athreed$}&
\multicolumn{1}{c}{$\Athreed-\Fthreed$}  &
\multicolumn{1}{c}{$\Bthreed$} &
\multicolumn{1}{l}{$\Bthreed-\Fthreed$}  &
\multicolumn{1}{c}{$\Cthreed$} &
\multicolumn{1}{c}{$\Cthreed-\Fthreed$}  &
\multicolumn{1}{r}{$\Fthreed$}  \\
\hline
\endhead
 ZPVE~~~~&   $0$& $0$ &  $A_{1}$ & 51.200&       0.000&      51.200&       0.000&      51.200&       0.000&      51.200\\    
1~~~~&   $1$&   $0$ &  $F_{2}$ &   9.107&      $-$0.002&       9.109&       0.000&       9.109&       0.000&       9.109\\
2~~~~&   $1$&   $0$ &  $F_{2}$ &  9.107&      $-$0.002&       9.109&       0.000&       9.109&       0.000&       9.109\\
3~~~~&   $1$&   $0$ &  $F_{2}$ &   9.109&       0.000&       9.109&       0.000&       9.109&       0.000&       9.109\\
4~~~~&   $0$&   $1$ &  $A_{1}$ &  29.188&       0.000&      29.188&       0.000&      29.188&       0.000&      29.188\\
5~~~~&   $2$&   $0$ &  $F_{2}$ & 31.384&      $-$0.004&      31.388&       0.000&      31.388&       0.000&      31.388\\
6~~~~&   $2$&   $0$ &  $F_{2}$ & 31.384&      $-$0.004&      31.388&       0.000&      31.388&       0.000&      31.388\\
7~~~~&   $2$&   $0$ &  $F_{2}$ &  31.388&       0.000&      31.388&       0.000&      31.388&       0.000&      31.388\\
8~~~~&   $2$&   $0$ &  $E$  & 31.942&       0.000&      31.942&       0.000&      31.942&       0.000&      31.942\\
9~~~~&   $2$&   $0$ &  $E$  & 31.942&       0.000&      31.942&       0.000&      31.942&       0.000&      31.942\\
10~~~~&   $1$&   $1$ &  $F_{2}$ &  44.570&      $-$0.004&      44.573&       0.000&      44.573&       0.000&      44.573\\
11~~~~&   $1$&   $1$ &  $F_{2}$ & 44.570&      $-$0.004&      44.573&       0.000&      44.573&       0.000&      44.573\\
12~~~~&   $1$&   $1$ &  $F_{2}$ &  44.573&       0.000&      44.573&       0.000&      44.573&       0.000&      44.573\\
13~~~~&   $0$&   $2$ &  $A_{1}$ & 53.036&       0.000&      53.036&       0.000&      53.036&       0.000&      53.036\\
14~~~~&   $2$&   $1$ &  $F_{2}$ & 56.228&      $-$0.004&      56.232&       0.000&      56.232&       0.000&      56.232\\
15~~~~&   $2$&   $1$ &  $F_{2}$ & 56.228&      $-$0.004&      56.232&       0.000&      56.232&       0.000&      56.232\\
16~~~~&   $2$&   $1$ &  $F_{2}$ & 56.232&       0.000&      56.232&       0.000&      56.232&       0.000&      56.232\\
17~~~~&   $2$&   $1$ &  $E$ & 64.046&       0.000&      64.046&       0.000&      64.046&       0.000&      64.046\\
18~~~~&   $2$&   $1$ &  $E$ & 64.046&       0.000&      64.046&       0.000&      64.046&       0.000&      64.046\\
19~~~~&   $3$&   $0$ &  $F_{2}$ &  65.825&      $-$0.013&      65.837&       0.000&      65.837&       0.000&      65.837\\
20~~~~&   $3$&   $0$ &  $F_{2}$ & 65.825&      $-$0.013&      65.837&       0.000&      65.837&       0.000&      65.837\\
21~~~~&   $3$&   $0$ &  $F_{2}$ & 65.837&       0.000&      65.837&       0.000&      65.837&       0.000&      65.837\\
22~~~~&   $1$&   $2$ &  $F_{1}$ & 66.066&      $-$0.004&      66.070&       0.000&      66.070&       0.000&      66.070\\
23~~~~&   $1$&   $2$ &  $F_{1}$ & 66.066&      $-$0.004&      66.070&       0.000&      66.070&       0.000&      66.070\\
24~~~~&   $1$&   $2$ &  $F_{1}$ & 66.070&       0.000&      66.070&       0.000&      66.070&       0.000&      66.070\\
25~~~~&   $0$&   $3$ &  $A_{1}$ & 70.313&       0.000&      70.313&       0.000&      70.313&       0.000&      70.313\\
26~~~~&   $3$&   $0$ &  $A_{1}$ & 73.497&       0.000&      73.497&       0.000&      73.497&       0.000&      73.497\\
27~~~~&   $2$&   $2$ &  $F_{2}$ & 75.340&      $-$0.007&      75.347&       0.000&      75.347&       0.000&      75.347\\
28~~~~&   $2$&   $2$ &  $F_{2}$ & 75.340&      $-$0.007&      75.347&       0.000&      75.347&       0.000&      75.347\\
29~~~~&   $2$&   $2$ &  $F_{2}$ & 75.347&       0.000&      75.347&       0.000&      75.347&       0.000&      75.347\\
30~~~~&   $1$&   $3$ &  $F_{2}$ & 80.280&      $-$0.003&      80.283&       0.000&      80.283&       0.000&      80.283\\
31~~~~&   $1$&   $3$ &  $F_{2}$ & 80.280&      $-$0.003&      80.283&       0.000&      80.283&       0.000&      80.283\\
32~~~~&   $1$&   $3$ &  $F_{2}$ & 80.283&       0.000&      80.283&       0.000&      80.283&       0.000&      80.283\\
33~~~~&   $0$&   $4$ &  $A_{1}$ & 83.085&       0.000&      83.085&       0.000&      83.085&       0.000&      83.085\\
34~~~~&   $1$&   $4$ &  $F_{2}$ & 88.186&      $-$0.003&      88.189&       0.000&      88.189&       0.000&      88.189\\
35~~~~&   $1$&   $4$ &  $F_{2}$ & 88.186&      $-$0.003&      88.189&       0.000&      88.189&       0.000&      88.189\\
36~~~~&   $1$&   $4$ &  $F_{2}$ & 88.189&       0.000&      88.189&       0.000&      88.189&       0.000&      88.189\\
37~~~~&   $2$&   $4$ &  $E$ & 88.826&       0.000&      88.826&       0.000&      88.826&       0.000&      88.826\\
38~~~~&   $2$&   $4$ &  $E$ & 88.826&       0.000&      88.826&       0.000&      88.826&       0.000&      88.826\\
39~~~~&   $0$&   $5$ &  $A_{1}$ & 89.427&       0.000&      89.427&       0.000&      89.427&       0.000&      89.427\\
 \hline
\end{longtable}
\end{center}

%%%%%%%%%%%%%%%%%%%%%%%%%%%%%%%%%%%%%%%%%%%%%%%%%%%%%%%%%%%%%%%%%%%%%%%%%%%%%%%%%%%%%%%%%%%%%%%%%%%%%%%%%%%%
\subsubsection{Intermolecular radial representation: Laguerre and Morse-tridiagonal basis sets}     
If we choose the $\mathcal{L}_{n}^{(\alpha)}$ generalized Laguerre basis functions (with $\alpha=2$) for
the $R$ radial coordinate, we have to use a large number, more than 30, basis functions to converge
the vibrational bound states.
Since in the present work we focus on the computation of bound states, 
it is better to use tridiagonal Morse basis set 
 \cite{doi:10.1063/1.463044,doi:10.1063/1.451775,doi:10.1063/1.444316}. 
The parameters of Morse function functions were determined according to the equations in Ref.~\cite{doi:10.1063/1.463044} with $D=143.49$ cm$^{-1}$, $\alpha=0.65$, and $\gamma=0.00033$. These parameters were adjusted to obtain 13 functions that recover the exact vibrational energies  for the bound states of the radial Hamiltonian
\begin{eqnarray}
  \hat{H}_{R}
  = 
  -\frac{1} {2\mu_{{\rm CH}_{4},{\rm Ar}}} \pd{^2}{R^2}
  +V(R,\theta^{\rm eq},\phi^{\rm eq})
\end{eqnarray}
where $\mu_{{\rm CH}_4,{\rm Ar}}$ is the reduced mass of methane and argon, 
and $\theta^{\rm eq}$ and $\phi^{\rm eq}$ are the equilibrium values of the 3D PES. 
Since the CH$_{4}\cdot$Ar complex is a very isotropic system, 
the parameters and the radial basis set determined in this way should be useful 
over the entire range of the $\theta\in[0,\pi]$ and $\phi\in[0,2\pi)$ coordinates.

%%%%%%%%%%%%%%%%%%%%%%%%%%%%%%%%%%%%%%%%%%%%%%%%%%%%%%%%%%%%%%%%%%%%%%%%%%%%%%%%%%%%%%%%%%%%%%%%%%%%%%%%%%%%    
\subsection{Analysis of the intramolecular representation: vibrational states of CH$_4$ \label{ch:intra}}
The vibrational basis set used to describe the intramolecular vibrational dynamics, 
\emph{i.e.,} vibrations of the methane molecule, 
was constructed from the harmonic oscillator basis set with 
the standard pruning condition, $\sum_{k=1}^9 n_{q_k} \le b$ in Eq.~(\ref{eq:ourbasispruning}),  
and the Smolyak quadrature with $\sum_{k=1}^9 g^{q_k}(i_{q_k}) \le H$ in Eq.~(\ref{eq:ourgridpruning}).
Table~\ref{ch4res} shows the convergence of the lowest vibrational states by increasing $b$ and $H$.

As to the 12D computation of CH$_4\cdot$Ar, the bound states correspond to 
the zero-point vibrational state (ZPV) of CH$_4$, we focused on the lowest-energy states of CH$_4$.
Of course, more accurate results for the isolated methane molecule can be obtained 
by increasing the size of the Smolyak grid, which is perfectly feasible for a 9D computation.

In a minimalistic setup (to be transferred for the 12D computations), we chose a representation 
which allowed us to converge the fundamental vibrational energies within 1~\cm.
In this representation the 9D Smolyak grid includes more than 100\,000 points, which 
is approximately an order of magnitude larger than what is necessary for a meaningful
representation of the zero-point vibration.

% \clearpage
\begin{center}
% ~\vspace{-2cm}
\begin{longtable}{@{}lcccccccc@{}}
\caption{%
Deviation of the vibrational energies, \cm,  of the CH$_4$ molecule
obtained with GENIUSH-Smolyak with a pruned basis and grid,
from the tightly converged results of Ref.~\cite{wang2014using},
with increasing the $b$ and $H$ values in the basis and 
the grid pruning conditions, Eqs.~(\ref{eq:ourbasispruning}) and (\ref{eq:ourgridpruning}),
respectively. 
In general, $H=b+D+2\geq b+D$ was found to be sufficient to converge the results 
(note that $D=9$ for isolated methane). 
The corresponding number of Smolyak points, $N_{\rm Smol}$, is also shown.
\label{ch4res}}\\
\hline\\[-0.7cm]\hline\\[-0.7cm]
\multicolumn{1}{l}{$n$} &
\multicolumn{7}{c}{Deviation from Ref.~\cite{wang2014using}} &
\multicolumn{1}{c}{Ref.~\cite{wang2014using}} \\
\cline{2-8}
\multicolumn{1}{r}{$b$:} &
\multicolumn{1}{c}{2} &
\multicolumn{1}{c}{3} &
\multicolumn{1}{c}{4} &
\multicolumn{1}{c}{5} &
\multicolumn{1}{c}{6} &
\multicolumn{1}{c}{7} &
\multicolumn{1}{c}{8} &
\multicolumn{1}{c}{} \\
\multicolumn{1}{r}{{$H_\text{9D}$:}} &
\multicolumn{1}{c}{13} &
\multicolumn{1}{c}{14} &
\multicolumn{1}{c}{15} &
\multicolumn{1}{c}{16} &
\multicolumn{1}{c}{17} &
\multicolumn{1}{c}{18} &
\multicolumn{1}{c}{19} &
\multicolumn{1}{c}{} \\
\multicolumn{1}{r}{$N_\text{Smol}$:} &
\multicolumn{1}{c}{3\ 481} &
\multicolumn{1}{c}{11\ 833} &
\multicolumn{1}{c}{35\ 929} &
\multicolumn{1}{c}{97\ 561} &
\multicolumn{1}{c}{241\ 201} &
\multicolumn{1}{c}{556\ 707} &
\multicolumn{1}{c}{1\ 202\ 691} \\
\hline
\endfirsthead
\endhead
 ZPV&       41.18&        2.51&        0.66&        0.57&        0.07&        0.02&        0.02&     9651.29\\
 1&       47.81&       44.46&        3.03&        0.81&        0.65&        0.09&        0.03&    10961.76\\
 2&       47.81&       44.45&        3.03&        0.81&        0.65&        0.09&        0.03&    10961.76\\
 3&       47.82&       44.45&        3.03&        0.81&        0.65&        0.09&        0.03&    10961.76\\
 4&       45.91&       42.46&        2.97&        0.75&        0.61&        0.08&        0.03&    11184.76\\
 5&       45.93&       42.47&        2.97&        0.75&        0.61&        0.08&        0.03&    11184.76\\
 6&       81.12&       57.35&       46.15&        4.79&        1.23&        0.74&        0.14&    12238.29\\
 7&       75.65&       55.48&       48.12&        4.18&        1.14&        0.77&        0.13&    12265.12\\
 8&       75.73&       55.48&       48.11&        4.17&        1.14&        0.77&        0.13&    12265.13\\
 9&       76.42&       55.48&       48.11&        4.17&        1.13&        0.77&        0.13&    12265.13\\
10&       65.82&       53.24&       47.43&        3.49&        1.00&        0.72&        0.11&    12275.73\\
11&       65.85&       53.24&       47.43&        3.49&        1.00&        0.72&        0.11&    12275.74\\
12&       78.12&       53.43&       43.26&        3.87&        1.03&        0.66&        0.11&    12481.49\\
13&       78.77&       53.48&       43.26&        3.88&        1.03&        0.66&        0.11&    12481.49\\
14&       78.77&       53.48&       43.27&        3.87&        1.03&        0.66&        0.11&    12481.49\\
15&       72.37&       51.38&       45.63&        3.57&        0.93&        0.69&        0.10&    12497.25\\
16&       72.37&       51.38&       45.65&        3.56&        0.94&        0.69&        0.10&    12497.25\\
17&       73.08&       51.44&       45.65&        3.57&        0.94&        0.69&        0.10&    12497.26\\
18&       83.68&       72.76&       15.82&        2.88&        1.76&        0.47&        0.12&    12568.47\\
19&       86.04&       74.55&       16.18&        2.89&        1.79&        0.48&        0.12&    12670.73\\
20&       86.07&       74.56&       16.18&        2.89&        1.79&        0.48&        0.12&    12670.73\\
21&       86.07&       74.56&       16.18&        2.89&        1.79&        0.48&        0.12&    12670.73\\
\hline\\[-0.7cm]\hline\\[-0.7cm]
\end{longtable}
\end{center}

%%%%%%%%%%%%%%%%%%%%%%%%%%%%%%%%%%%%%%%%%%%%%%%%%%%%%%%%%%%%%%%%%%%%%%%%%%%%%%%%%%%%%%%%%%%%%%%%%%%%%%%%%%%%
\section{Full-dimensional (12D) results for methane-argon \label{ch:12dim}}
All bound-state vibrational energies were computed for the CH$_{4}\cdot$Ar complex 
in full (12D) vibrational dimensionality (Table \ref{12D}). 
The basis and the grid representations are selected based on 
the convergence tests carried out for the inter- and intra-molecular 
representations (Sections~\ref{ch:inter} and \ref{ch:intra}).
Concerning the intermolecular representation, it is composed of 
Morse-tridiagonal basis functions with $N^{\rm max}_{R}=12$, 
sincot-Legendre-DVR basis functions with $N^{\rm max}_{c}=20$, and 
Fourier functions with $N^{\rm max}_{\phi}=17$.
The number of quadrature points was
$K_{R}^{\rm max}=15$, $K_{c}^{\rm max}=21$, and $K_{\phi}^{\rm max}=20$.
As to the methane fragment, 
we used four different intramolecular representations, 
with $b=0,1,2,$ and 3 values, 
which allowed us to check the convergence of 
the ZPVE and the vibrational energies in the full-dimensional treatment.

Table~\ref{tab:metdef12d} gives an overview of the orders of magnitudes of 
the basis and the grid representations employed in the final 12D computations.
The largest computation (set $D$ in the table) includes
82\,002\,690 (8.20$\cdot 10^7$) quadrature points and 1\,021\,020 ($1.02\cdot 10^6$) 
basis functions.
The numerical KEO terms, Eq.~(\ref{H12D}), and the PES are stored 
as double precision reals (in Fortran) at every grid
point, which amounts to a $(12\cdot 13/2 + 12+1) \cdot 8.20 \cdot 10^7=60$~GB
memory usage. The dimensionality of the Lanczos vectors are determined by the number
of basis functions, so one Lanczos vector occupies a negligible amount of 8~MB of memory.
To multiply a trial vector with the Hamiltonian matrix took
ca. 230 seconds on 51 processor cores,
and we had to perform ca.~10\,000 matrix-vector multiplications
to obtain the 40 states reported in Table~\ref{12D} using an in-house Lanczos
implementation (it might be possible to reduce the number of matrix-vector products
with a Lanczos and a pre-conditioning algorithm optimized for the present system).

Based on the isolated-methane test computations (Table~\ref{ch4res})
the error in the ZPVE for $b=2$ and $3$ is 41 and 2.5~\cm, respectively.
The vibrational energies of the complex (referenced to the ZPVE) change less than 0.01~\cm\
by increasing the $b$ value from 2 to 3, hence we may accept them as converged for 
$b=3$. 
The ZPVE of the complex is probably accurate within a few \cm\ with $b=3$ 
similarly to the case of the isolated methane (Table~\ref{ch4res}).
We only note that a full 12D computation with $b=4$ would also be feasible
with the current implementation, but it would only change the ZPVE, since the 
vibrational energies were converged already with $b=3$.

We also show the $b=0$ results, which correspond to 
a single harmonic oscillator function for methane 
(the product of the zeroth harmonic oscillator basis functions for $q_1,\ldots,q_9$).
Since the present model includes only kinetic coupling 
(the PES coupling is also probably very small), the deviation of $\tilde\nu_A$ ($b=0$)
and $\tilde\nu_A$ ($b=3$) is due to the structural differences of methane:
the effective structure for the $b=0$ ground-state harmonic oscillator basis function 
is the equilibrium structure, whereas $b>0$ accounts for structural distortions 
due to anharmonicity effects.  This change is related to the common wisdom in 
reduced-dimensionality computations of weakly-bound complexes that it is
better to use effective (vibrationally averaged) monomer structures than equilibrium monomer structures \cite{JeJaSzJe00}.
In agreement with this prescription, the 3D computation (column $\Fthreed$ 
in Table~\ref{tab:metdef12d}) performed with an effective methane structure 
corresponding to the (isolated) ground-state vibration very well reproduces the 12D result
(remember that only kinetic coupling is included in the present computation, due to the lack
of a 12D fully coupled PES).

\begin{table}
 \caption{%
   Intramolecular (methane, ``Met'') basis set and grid choices used in 
   the 12D CH$_4\cdot$Ar vibrational computations
   with the basis and grid pruning conditions $n_{q_{1}}+\ldots+n_{q_{9}}\le b$  and
   $i_{q_{1}}+\ldots+i_{q_{9}}\le H$, respectively. 
   $H=b+D+2\geq b+D$ was found to be sufficient to converge the results ($D=12$).
   The number of basis functions, $N$, and grid points, $K$, are also given 
   for the methane (``Met'', ``Smol''), for the intermolecular (``Inter''),
   and for the full (12D) computations.
  \label{tab:metdef12d}
}
\begin{tabular}{@{}c@{\ \ \ }c@{\ \ \ }c @{\ \ \ \ }c @{\ \ \ \ }c cc ccc c @{}}
  \hline\hline\\[-0.7cm]
  & \multicolumn{4}{c}{Intramolecular (CH$_4$, 9D)}              
  && \multicolumn{2}{c}{Intermolecular (3D)} 
  && \multicolumn{2}{c}{CH$_4\cdot$Ar (12D)}  \\
  \cline{2-5} \cline{7-8} \cline{10-11} \\[-0.7cm]
  Label & $b$ & $H$ & 
  $N_{\rm Met}$ & $K_{\rm Smol} / 10^3$ && 
  $N_{\rm Inter}/10^3$ & $K_{\rm Inter}/10^3$  && 
  $N_{\rm 12D}/10^5$  & $K_{\rm 12D}/10^7$ \\
  \hline\\[-0.7cm]
  $A$     &  0 & 14 & 1   & $0.163$ && 4.28 & 6.30 && 0.0464 & 0.113 \\
  $B$     &  1 & 15 & 10  & $0.871$ && 4.28 & 6.30 && 0.464 & 0.604 \\
  $C$     &  2 & 16 & 55  & $3.48$  && 4.28 & 6.30 && 2.55 & 2.41 \\
  $D$     &  3 & 17 & 220 & $11.8$  && 4.28 & 6.30 && 10.2 & 8.20 \\
  \hline\hline\\[-0.7cm]
\end{tabular}
\end{table}

% \clearpage
\begin{center}
\begin{longtable}{@{}r@{\ \ \ }rr@{\ \ \ \ \ }rr@{\ \ \ \ \ }rr@{\ \ \ \ \ } rr@{}}
\caption{Vibrational bound-state energies, $\tilde\nu$ in \cm, referenced to the ZPVE 
of CH$_{4}\cdot$Ar computed in full (12D) vibrational dimensionality
using the GENIUSH program extended with the Smolyak algorithm in the present work. 
The potential energy was approximated with the sum of the molecule-atom interaction 
PES of Ref.~\cite{B009741L,doi:10.1063/1.1506153} 
and 
the isolated methane PES of Ref.~\cite{wang2014using}.
The $A,B,C,$ and $D$ basis and grid representations defined in Table~\ref{tab:metdef12d} 
correspond to an increasing $b=0,1,2,$ and 3 value in 
the methane basis functions' pruning condition, Eq.~(\ref{eq:ourbasispruning}).
Convergence of the results can be estimated based on 
the deviation from computation $D$,  $\Delta\tilde\nu_i=\tilde\nu_i-\tilde\nu_D$ (see also Table~\ref{ch4res}).
For comparison, the benchmark 3D computation with a fixed, 
effective ($v=0$) methane geometry ($\Fthreed$) is also shown (taken from Table~\ref{1table}).
\label{12D} } \\
\hline\hline \\[-0.7cm]
\multicolumn{1}{c}{} &
\multicolumn{7}{c}{12D} &
\multicolumn{1}{r}{3D} \\
\cline{2-8}\\
\multicolumn{1}{c}{Label} &
\multicolumn{2}{c}{$A$ ($b=0$)} &
\multicolumn{2}{c}{$B$ ($b=1$)} &
\multicolumn{2}{c}{$C$ ($b=2$)} &
\multicolumn{1}{c}{$D$ ($b=3$)} &
\multicolumn{1}{r}{(Table~\ref{1table})} \\
\multicolumn{1}{c}{} &
\multicolumn{1}{c}{$\tilde\nu_A$} &
\multicolumn{1}{c}{$\Delta\tilde\nu_A$} &
\multicolumn{1}{c}{$\tilde\nu_B$} &
\multicolumn{1}{c}{$\Delta\tilde\nu_B$} &
\multicolumn{1}{c}{$\tilde\nu_C$} &
\multicolumn{1}{c}{$\Delta\tilde\nu_C$} &
\multicolumn{1}{c}{$\tilde\nu_D$} &
\multicolumn{1}{r}{$\tilde\nu_{\Fthreed}$} \\
\hline
\endfirsthead
\multicolumn{5}{c}{{\tablename} \thetable{}  Continued} \\[0.5ex]
\multicolumn{1}{c}{} &
\multicolumn{1}{c}{$\tilde\nu_A$} &
\multicolumn{1}{c}{$\Delta\tilde\nu_A$} &
\multicolumn{1}{c}{$\tilde\nu_B$} &
\multicolumn{1}{c}{$\Delta\tilde\nu_B$} &
\multicolumn{1}{c}{$\tilde\nu_C$} &
\multicolumn{1}{c}{$\Delta\tilde\nu_C$} &
\multicolumn{1}{c}{$\tilde\nu_D$} &
\multicolumn{1}{r}{$\tilde\nu_{\Fthreed}$} \\
\hline
\endhead
 ZPV&      9695.262&     132.242&    9604.164&      41.144&    9600.706&      37.686&    9563.019       &      51.200\\
 1&       9.398&       0.285&       9.139&       0.026&       9.112&      $-$0.002&       9.113       &       9.109\\
 2&       9.398&       0.285&       9.139&       0.026&       9.112&      $-$0.002&       9.113       &       9.109\\
 3&       9.398&       0.285&       9.139&       0.026&       9.112&      $-$0.002&       9.113       &       9.109\\
 4&      29.275&       0.086&      29.197&       0.008&      29.189&      $-$0.000&      29.189       &      29.188\\
 5&      31.970&       0.575&      31.447&       0.052&      31.392&      $-$0.003&      31.395       &      31.388\\
 6&      31.970&       0.575&      31.447&       0.052&      31.392&      $-$0.003&      31.395       &      31.388\\
 7&      31.970&       0.574&      31.447&       0.052&      31.392&      $-$0.003&      31.395       &      31.388\\
 8&      32.687&       0.736&      32.016&       0.066&      31.946&      $-$0.004&      31.950       &      31.942\\
 9&      32.687&       0.736&      32.017&       0.066&      31.946&      $-$0.004&      31.950       &      31.942\\
10&      45.042&       0.463&      44.620&       0.041&      44.576&      $-$0.003&      44.579       &      44.573\\
11&      45.042&       0.463&      44.620&       0.041&      44.577&      $-$0.003&      44.579       &      44.573\\
12&      45.042&       0.463&      44.620&       0.041&      44.577&      $-$0.003&      44.579       &      44.573\\
13&      53.156&       0.119&      53.048&       0.011&      53.036&      $-$0.001&      53.037       &      53.036\\
14&      57.039&       0.799&      56.313&       0.073&      56.236&      $-$0.005&      56.240       &      56.232\\
15&      57.039&       0.799&      56.313&       0.073&      56.236&      $-$0.005&      56.240       &      56.232\\
16&      57.039&       0.799&      56.313&       0.073&      56.236&      $-$0.005&      56.240       &      56.232\\
17&      64.807&       0.753&      64.122&       0.068&      64.050&      $-$0.004&      64.055       &      64.046\\
18&      64.807&       0.753&      64.122&       0.068&      64.050&      $-$0.004&      64.055       &     64.046\\
19&      66.819&       0.970&      65.989&       0.141&      65.839&      $-$0.009&      65.848       &      65.837\\
20&      66.819&       0.971&      65.989&       0.141&      65.839&      $-$0.009&      65.848       &      65.837\\
21&      66.819&       0.970&      65.989&       0.141&      65.839&      $-$0.009&      65.848       &      65.837\\
22&      67.414&       1.337&      66.143&       0.066&      66.072&      $-$0.004&      66.076       &      66.070\\
23&      67.414&       1.337&      66.143&       0.067&      66.072&      $-$0.004&      66.076       &      66.070\\
24&      67.414&       1.337&      66.143&       0.067&      66.072&      $-$0.004&      66.077       &      66.070\\
25&      70.705&       0.388&      70.360&       0.043&      70.314&      $-$0.003&      70.317       &     70.313\\
26&      74.623&       1.118&      73.597&       0.092&      73.499&      $-$0.006&      73.505       &     73.497\\
27&      76.276&       0.920&      75.438&       0.081&      75.351&      $-$0.005&      75.356       &      75.347\\
28&      76.276&       0.920&      75.438&       0.081&      75.351&      $-$0.005&      75.356       &      75.347\\
29&      76.276&       0.920&      75.442&       0.086&      75.351&      $-$0.005&      75.356       &     75.347\\
30&      80.808&       0.517&      80.338&       0.047&      80.287&      $-$0.003&      80.290       &      80.283\\
31&      80.808&       0.517&      80.338&       0.047&      80.288&      $-$0.003&      80.291       &      80.283\\
32&      80.808&       0.517&      80.337&       0.047&      80.288&      $-$0.003&      80.291       &     80.283\\
33&      83.156&       0.067&      83.093&       0.004&      83.088&      $-$0.000&      83.088       &     83.085\\
34&      88.844&       0.647&      88.255&       0.057&      88.194&      $-$0.004&      88.197       &     88.189\\
35&      88.844&       0.647&      88.254&       0.056&      88.194&      $-$0.004&      88.198       &     88.189\\
36&      88.844&       0.647&      88.254&       0.056&      88.194&      $-$0.004&      88.198       &     88.189\\
37&      89.588&       0.753&      88.903&       0.068&      88.830&      $-$0.004&      88.835       &     88.826\\
38&      89.588&       0.753&      88.902&       0.067&      88.830&      $-$0.004&      88.835       &     88.826\\
39&      89.505&       0.017&      89.431&      $-$0.057&      89.488&       0.000&      89.488       &     89.427\\
\hline\hline \\[-0.7cm]
\end{longtable}          
\end{center}

%%%%%%%%%%%%%%%%%%%%%%%%%%%%%%%%%%%%%%%%%%%%%%%%%%%%%%%%%%%%%%%%%%%%%%%%%%%%%%%%%%%%%%%
%
% Summary, conclusion, and outlook
%
%%%%%%%%%%%%%%%%%%%%%%%%%%%%%%%%%%%%%%%%%%%%%%%%%%%%%%%%%%%%%%%%%%%%%%%%%%%%%%%%%%%%%%%
% \clearpage
\section{Summary, conclusions, and outlook \label{ch:sum}}
The numerical kinetic-energy operator (KEO) approach as implemented in the 
GENIUSH program \cite{MaCzCs09} has been extended with 
the Smolyak algorithm \cite{tc-gab1,tc-gab2}, which opens a promising route
towards variational (ro)vibrational computations for polyatomic systems with multiple 
large-amplitude motions.

A direct, variational solution of the (ro)vibrational Schr\"odinger equation
of polyatomic systems (without imposing constraints on the coordinates)
is difficult due to the high vibrational dimensionality, which generates an exponential growth in
the direct-product basis used to represent the wave functions,
and an exponential growth in the direct-product grid necessary to calculate integrals of multi-dimensional
operator terms in the Hamiltonian.

If coordinates well-suited for the motions in the system and 
good zeroth-order basis functions can be found for each coordinate, 
it is not necessary to use a direct-product basis, but a much smaller, `pruned' basis 
can be defined, the size of which does not scale exponentially
with the number of vibrational degrees of freedom.
If it is possible to prune a direct-product basis, 
it is also possible to find a pruned product grid to calculate integrals.
The Smolyak scheme of Avila and Carrington \cite{tc-gab1,tc-gab2} makes it possible
to define non-product (Smolyak) grids, which are orders of magnitude smaller
than a direct-product grid but which retain some of the practical features of
a direct-product grid. Most importantly, Smolyak grids can be efficiently used in
computing matrix-vector products and efficient matrix-vector products allow us
to compute eigenvalues and eigenfunctions with an iterative (Lanczos) eigensolver
without storing or even explicitly computing the Hamiltonian matrix elements.

In the present work, the combination of these ideas with the numerical KEO approach of GENIUSH
were elaborated and explained for all stages of the vibrational computation of
the floppy CH$_4\cdot$Ar complex treated in full vibrational dimensionality. 
Due to the highly fluxional nature of this system,
regions of the curvilinear coordinate domains above which the KEO
has singularities are dynamically important.

In a fully finite basis representation (FBR) treatment of the numerical KEO, the Hamiltonian matrix fails to be 
Hermitian due to inaccurate integration of the singularities in general coordinates. 
As a practical way to avoid these singularity problems in FBR, 
we proposed to use (efficient) DVRs and an inherently symmetric form of the general
KEO for the singular coordinate(s), which ensures a symmetric matrix representation by construction
and correct limiting (convergence) behavior at the same time.
In practice, this hybrid DVR-FBR treatment allows us to
converge all bound vibrational states of CH$_4\cdot$Ar.

In general, this hybrid DVR-FBR approach makes it possible to continue using 
1) numerical KEOs; and 
2) a general and simple starting product basis sets and grids 
(both pruned according to physically motivated restrictions)
for systems with multiple large-amplitude motions; 
and ultimately, to (further) develop a universal, black-box-type 
(ro)vibrational procedure practical for polyatomic systems.
Extension of the algorithm for $J>0$ rotational quantum number is straightforward, 
limitations might be set by the memory requirements and the computational time.

We can foresee future possible improvements of the present procedure
to (at least partially) eliminate the current bottlenecks in terms of memory usage 
(storage of the numerical KEO terms over the grid, see for example Ref.~\cite{NaLa18}) and 
perhaps also in terms of the computational cost.
Furthermore, the present developments, in particular the fact that the
Smolyak grid is several orders of magnitude smaller than the direct product grid, 
can be combined with the basis-set contraction idea \cite{CM1,YuBaYaThJeTe09,WaCa18WW}. 
With these or other developments, it will become possible to directly access 
the predissociation spectral range corresponding to the molecule's fundamental 
(and lowest overtone) vibrations in weakly or more strongly bound complexes of 
the size of CH$_4\cdot$Ar, \emph{i.e.,} with $D=12$ or perhaps beyond this value.
In general, a careful choice of the coordinate set, the basis, and the grid representation 
are required to make full use of the ideas combined, developed, and described in the present
work. We hope that these ideas will find applications, also beyond the realm of molecular complexes,
among high-dimensional molecular systems with multiple large-amplitude motions.

\clearpage
\vspace{1cm}
\noindent\textbf{Supplementary Material}\\
Definition of the normal coordinates used for the methane fragment is provided 
in the Supplementary Material.

\vspace{1cm}
\noindent\textbf{Acknowledgment}\\ 
Financial support of the Swiss National Science Foundation through
a PROMYS Grant (no. IZ11Z0\_166525) is gratefully acknowledged. 
We thank PRACE for a Preparatory Access Grant during 2017--2018,
and NIIFI for allocating computer time 
at The Hungarian Computing Infrastructure (Miskolc node).
Xiao-Gang Wang and Tucker Carrington 
helped us to get through a difficult stage of the development work 
by sharing their well-tested sincot-Legendre-DVR implementation (codvr.f90) with us \cite{WaCa12}. 
We are indebted to
our colleagues and co-authors, in particular
Tucker Carrington, Attila Cs\'asz\'ar, and G\'abor Czak\'o, 
with whom we worked together on Refs.~\cite{tc-gab1} and \cite{MaCzCs09},
and also the colleagues who later contributed to 
the further developments and successful applications 
of the initial methods over the past 10 years.

% \bibliography{paper.bib}           

\end{document}